\documentclass[english,aps,a4paper,prd,superscriptaddress,preprintnumbers,nofootinbib,preprint,floatfix,showpacs,amssymb]{revtex4}
\usepackage{amsmath}
\usepackage{url}
\usepackage{xcolor}
\usepackage{hyperref}
\hypersetup{colorlinks=true,linkcolor=redLinks,citecolor=greenLinks,urlcolor=redLinks,
pdfborder={0 0 1}}
\hypersetup{
    bookmarks=true,         
    unicode=false,          
    pdftoolbar=true,        
    pdfmenubar=true,        
    pdffitwindow=false,     
    pdfstartview={FitH},    
    pdftitle={My title},    
    pdfauthor={Author},     
    pdfsubject={Subject},   
    pdfcreator={Creator},   
    pdfproducer={Producer}, 
    pdfkeywords={keyword1, key2, key3}, 
    pdfnewwindow=true,      
    colorlinks=false,       
    linkcolor=red,          
    citecolor=green,        
    filecolor=magenta,      
    urlcolor=cyan           
}
\colorlet{shadecolor}{gray!15}
\definecolor{greenLinks}{rgb}{0, 0.6, 0} 
\definecolor{blueLinks}{rgb}{0, 0, 0.6}
\definecolor{redLinks}{rgb}{0.6, 0, 0}
\definecolor{tempText}{rgb}{0.55, 0.10,0.67}
\definecolor{eprintLinks}{rgb}{0.4, 0.4, 0.4}
\definecolor{journalLinks}{rgb}{0.6, 0, 0}
\newcommand{\MYhref}[3][redLinks]{\href{#2}{\color{#1}{#3}}}%
\usepackage{subfig}
\usepackage{framed}
\usepackage{rotating}
\usepackage{slashed}     
\providecommand{\tabularnewline}{\\}
\usepackage{multirow} 
\usepackage[T1]{fontenc} 
\usepackage[utf8]{inputenc}
\usepackage{lmodern}
\usepackage{color,soul}
\makeatletter
\providecommand{\tabularnewline}{\\}
\usepackage{booktabs}
\usepackage{mathrsfs}   
\usepackage{units}
\usepackage{wasysym}
\def\lnv{lepton number violation }

\usepackage{float}
\newcommand{\bea}{\begin{eqnarray}}
\newcommand{\eea}{\end{eqnarray}}
\def\SM{$\mathrm{SU(3)_c \otimes SU(2)_L \otimes U(1)_Y}$ }
\def\ba{$$\begin{array}}
\def\ea{\end{array}$$}
\def\bra{$\begin{array}}
 \def\era{\end{array}$}
 
\newcommand{\sm}{Standard Model }
\newcommand{\g}{\,\mbox{GeV}}
\newcommand{\gmn}{\,g_{\mu\nu}}
\newcommand{\ppm}{\,(p_{1}-p_{2})_{\mu}}

\newcommand{\zg}{Z\gamma}
\newcommand{\gaga}{\gamma\gamma}
\newcommand{\hpm}{H^{\pm}}

\newcommand{\ddpm}{\Delta^{\pm\pm}}

\makeatother

\usepackage{babel}

\newcommand {\be} {\begin{equation}}
\newcommand {\ee} {\end{equation}}

\def\vev#1{\left\langle #1\right\rangle}

\newcommand{\nn}{\nonumber}

\newcommand{\AddrAHEP}{
  {\it AHEP Group, Instituto de F\'{\i}sica Corpuscular --
    C.S.I.C./Universitat de Val{\`e}ncia \\
    Edificio de Institutos de Paterna,
 C/Catedratico Jos\'e Beltr\'an, 2 E-46980 Paterna (Val\`{e}ncia) - SPAIN}}

\newcommand{\AddrLisb}{%
 Departamento de F\'\i sica and CFTP, Instituto Superior T\'ecnico\\
 Universidade de Lisboa, 
          Av. Rovisco Pais 1, 1049-001 Lisboa, Portugal }

\def\gsim{\raise0.3ex\hbox{$\;>$\kern-0.75em\raise-1.1ex\hbox{$\sim\;$}}}
\def\lsim{\raise0.3ex\hbox{$\;<$\kern-0.75em\raise-1.1ex\hbox{$\sim\;$}}}

\begin{document}


\title{Electroweak breaking and neutrino mass: \\
       ``invisible'' Higgs decays at the LHC \\ (Type II seesaw)}

\author{Cesar Bonilla} \email{cesar.bonilla@ific.uv.es}
\affiliation{\AddrAHEP}
\author{Jorge C. Rom\~ao}\email{jorge.romao@tecnico.ulisboa.pt}
\affiliation{\AddrLisb} 
\author{Jos\'e W. F. Valle} \email{valle@ific.uv.es}
\affiliation{\AddrAHEP} 


\begin{abstract}

  Neutrino mass generation through the Higgs mechanism not only
  suggests the need to reconsider the physics of electroweak symmetry
  breaking from a new perspective, but also provides a new
  theoretically consistent and experimentally viable paradigm.
  We illustrate this by describing the main features of the
  electroweak symmetry breaking sector of the simplest type-II seesaw
  model with spontaneous breaking of lepton number. After reviewing
  the relevant ``theoretical'' and astrophysical restrictions on the
  Higgs sector, we perform an analysis of the sensitivities of Higgs
  boson searches at the ongoing ATLAS and CMS experiments at the LHC,
  including not only the new contributions to the decay channels
  present in the \sm (SM) but also genuinely non-SM Higgs boson
  decays, such as ``invisible'' Higgs boson decays to majorons. We
  find sensitivities that are likely to be reached at the upcoming Run
  of the experiments.

\end{abstract}

\maketitle

\section{Introduction}

The electroweak breaking sector is a fundamental ingredient of the \sm
many of whose detailed properties remain open, even after the historic
discovery of the Higgs boson~\cite{Aad:2012tfa,Aad:2014eva}. The
electroweak breaking sector is subject to many restrictions following
from direct experimental searches at
colliders~\cite{Khachatryan:2014jba,Aad:2015gba}, as well as global
fits \cite{Baak:2012kk,Baak:2014ora} of precision
observables~\cite{PhysRevLett.65.964, ALTARELLI1991161,
  PhysRevD.46.381}. 
Moreover, its properties are may also be restricted by theoretical
consistency arguments, such as naturalness, perturbativity and
stability~\cite{Djouadi:2005gi}. The latter have long provided strong
motivation for extensions of the \sm such as those based on the idea
of supersymmetry.

Following the approach recently suggested in
Refs.~\cite{Bonilla:2015kna,Bonilla:2015eha} we propose to take
seriously the hints from the neutrino mass generation scenario to the
structure of the scalar sector. In particular, the most accepted
scenario of neutrino mass generation associates the smallness of
neutrino mass to their charge neutrality which suggests them to be of
Majorana nature due to some, currently unknown, mechanism of \lnv. The
latter requires an extension of the \SM Higgs sector and hence the
need to reconsider the physics of symmetry breaking from a new
perspective. In broad terms this would provide an alternative to
supersymmetry as paradigm of electroweak breaking.
Amongst its other characteristic features is the presence of doubly
charged scalar bosons, compressed mass spectra of heavy scalars
dictated by stability and perturbativity and the presence of
``invisible'' decays of Higgs bosons to the Nambu-Goldstone boson
associated to spontaneous \lnv and neutrino mass
generation~\cite{Bonilla:2015uwa}.

In this paper we study the invisible decays of the Higgs bosons in the
context of a type-II seesaw majoron model~\cite{Schechter:1981cv} in
which the neutrino mass is generated after spontaneous violation of
lepton number at some low energy scale, $\Lambda_\text{EW} \lesssim
\Lambda \sim
\mathcal{O}(\text{TeV})$~\cite{Joshipura:1992hp,Diaz:1998zg}~\footnote{The
  idea of the Majoron was first proposed in~\cite{Chikashige:1980ui}
  though in the framework of the Type I seesaw, not relevant for our
  current paper. On the other hand the triplet Majoron was suggested
  in~\cite{Gelmini:1980re} but has been ruled out since the first
  measurements of the invisible Z width by the LEP
  experiments. Regarding the idea of invisible Higgs decays was first
  given in Ref.~\cite{Shrock:1982kd}, though the early scenarios have
  been ruled out.}. This scheme requires the
presence of two lepton number--carrying scalar multiplets in the
extended \SM model, a singlet $\sigma$ and a triplet $\Delta$ under
SU(2) -- this seesaw scheme was called ``123''-seesaw model
in~\cite{Schechter:1981cv} and here we take the ``pure'' version of
this scheme, without right-handed neutrinos. The presence of the new
scalars implies the existence of new contributions to ``visible'' SM
Higgs decays, such as the $h\to\gamma\gamma$ decay channel, in
addition to intrinsically new Higgs decay channels involving the
emission of majorons, such as the ``invisible'' decays of the CP-even
scalar bosons. As a result, one can set upper limits on the invisible
decay channel based on the available data which restrict the
``visible'' channels.

The plan of this paper is as follows.  In the next section we describe
the main features of the symmetry breaking sector of the ``123'' type
II seesaw model. In section III we discuss the ``theoretical'' and
astrophysical constraints relevant for the Higgs sector. Taking these
into account, we study the sensitivities of Higgs boson searches at
the LHC to \sm scalar boson decays in section IV. Section V addresses
the non-SM Higgs decays of the model.  Section VI summarizes our
results and we conclude in section VII.

\section{The type-II seesaw model}                                                

Our basic framework is the ``123'' seesaw scheme originally proposed
in Ref.~\cite{Schechter:1981cv} whose Higgs sector contains, in
addition to the \SM scalar doublet $\Phi$, two lepton-number-carrying
scalars: a complex singlet $\sigma$ and a triplet $\Delta$.  All these
fields develop non-zero vacuum expectation values (vevs) leading to
the breaking of the \sm (SM) gauge group as well as the global
symmetry $U(1)_L$ associated to lepton number. The latter breaking
accounts for generation of the small neutrino masses.
Therefore, the scalar sector is given by 
\bea 
\Phi =\left[
\begin{array}{c}
\phi^{0}      \\
\phi^{-}\\ 
\end{array}\right]
\ \ \text{and}\ \
\Delta =\left[
\begin{array}{cc}
\Delta^{0}                   & \frac{\Delta^{+}}{\sqrt{2}} \\
\frac{\Delta^{+}}{\sqrt{2}}  & \Delta^{++}
\end{array}\right]
\eea 
with $L=0$ and $L=-2$, respectively, and the scalar field $\sigma$
with lepton number $L=2$. Below we will consider the required vev
hierarchies in the model.

\subsection{Yukawa Sector}

\begin{figure}[H]
  \centering
  \includegraphics[width=0.28\textwidth]{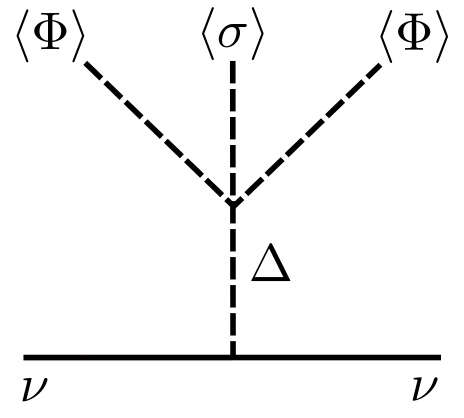} \quad
\vspace{-5pt}
  \caption{Diagram that generates non-zero neutrino mass in the 
  model.\label{numass}}
\end{figure}

Here we consider the simplest version of the seesaw scheme proposed in
Ref.~\cite{Schechter:1981cv} in which no right-handed neutrinos are
added, and only the \SM electroweak breaking sector is extended so as
to spontaneously break lepton number giving mass to neutrinos. Such
``123'' majoron--seesaw model is described by the \SM $\otimes~\rm
U(1)_{L}$ invariant Yukawa Lagrangian,
 \begin{equation}
  \mathcal{L}_Y = y^d_{ij}\overline{Q}_{i}u_{R_{j}}\Phi+y^u_{ij}\overline{Q}_{i}
                d_{R_{j}}\tilde{\Phi}+y^\ell_{ij} \overline{L_{i}}\ell_{R_{j}}\Phi+
                y^\nu_{ij} L^T_i C \Delta L_j + \text{h.c}.
 \end{equation}
In this model the neutrino mass (see Fig.~\ref{numass}) is given by,
\begin{equation}
 \label{mnu}
  m_\nu=y^{\nu} \kappa v_{1} \frac{v_2^2}{m_{\Delta}^2}  
 \end{equation}
 where $v_1$ and $v_2$ are the vevs of the singlet and the doublet,
 respectively. Here $\kappa$ is a dimensionless parameter that
 describes the interaction amongst the three scalar fields (see
 below), and $m_{\Delta}$ is the mass of the scalar triplet $\Delta$.

At this point we note that the smallness of neutrino mass
i.e. $$m_\nu\lsim 1~eV$$ may define interesting regions of the
parameter space in any neutrino mass generation model where the new
physics is expected to be hidden from direct observation.  In
particular, we are interested in spotting those regions accessible at
collider searches such as the ongoing experiments at the LHC (see
Ref.~\cite{Boucenna:2014zba} and references therein).

In our pure type II seesaw model where lepton number is spontaneously
violated at some low energy scale we have
$$m_{\nu}=y^\nu \vev{\Delta}$$ with the effective vev is given as 
$\vev{\Delta}=\mu \vev{ \Phi }^2/M_{\Delta}^2$ where $\Delta$ is the
isotriplet lepton--number--carrying scalar.  Here $\vev{\Phi}$ is
fixed by the mass of the $W$ boson and $$\mu=\kappa v_{1}$$ is the
dimensionful parameter responsible of lepton number violation, see
eq.~(\ref{mnu}). Therefore if $y^\nu\sim\mathcal{O}(1)$ and the mass
$M_{\Delta}$ lies at $1$~TeV region then one has that
$\vev{\Delta}\sim m_{\nu}$ and $\mu\sim 1$~eV.
Note that one may consider two situations: $v_{1}\gg \Lambda_{EW}$
(high-scale seesaw mechanism) in whose case the scalar singlet and the
invisible decays of the Higgs are decoupled~\cite{Joshipura:1992hp};
the second interesting case is when $\Lambda_{EW} \lesssim v_1
\lesssim$~few TeV (low-scale seesaw mechanism). In this case the
parameter $\kappa$ is the range $[10^{-14},10^{-16}]$ for
$y^\nu\sim\mathcal{O}(1)$.  In this case one has new physics at the
TeV region including the   ``invisible'' decays of the Higgs bosons.

Therefore, led by the smallness of the neutrino mass we can
qualitatively determine that the analysis to be carried out 
is characterized by having a vev hierarchy 
$$v_{1}\gtrsim v_2 \gg v_3$$ 
and the smallness of the coupling $\kappa$, that is $\kappa\ll1$.

\subsection{The scalar potential}

The scalar potential invariant under the \SM $\otimes~\rm U(1)_{L}$ 
symmetry is given by~\cite{Joshipura:1992hp,Diaz:1998zg}
\footnote{From now on we follow the notation and conventions
used in Ref.~\cite{Diaz:1998zg}.}
\bea
\label{eq:2}
V&=&\mu_1^{2}\sigma^{\ast}\sigma +\mu_{2}^2\Phi^{\dagger}\Phi
 +\mu_{3}^{2}\text{tr}(\Delta^{\dagger}\Delta)
 +\lambda_{1}(\Phi^{\dagger}\Phi)^2
 +\lambda_{2}[\text{tr}(\Delta^{\dagger}\Delta)]^2\nn\\
&+&\lambda_{3} \Phi^{\dagger}\Phi\text{tr}(\Delta^{\dagger}\Delta)
 +\lambda_4 \text{tr}(\Delta^{\dagger}\Delta\Delta^{\dagger}\Delta)
 +\lambda_5 (\Phi^{\dagger}\Delta^{\dagger}\Delta\Phi)
 +\beta_1 (\sigma^{\ast}\sigma)^2\nn\\
&+&\beta_2(\Phi^{\dagger}\Phi)(\sigma^{\ast}\sigma)
 +\beta_3 \text{tr}(\Delta^{\dagger}\Delta)
  (\sigma^{\ast}\sigma)-\kappa(\Phi^{T}\Delta\Phi\sigma +\text{h.c.}).
\eea
As mentioned above the scalar fields $\sigma$, $\phi$ and $\Delta$
acquire non-zero vacuum expectation values, $v_{1}$, $v_{2}$ and
$v_{3}$, respectively, so that, they can be shifted as follows,
\bea
&&\sigma=\frac{v_1}{\sqrt{2}}+\frac{R_1 +iI_1}{\sqrt{2}},\nn\\
&&\phi^{0}=\frac{v_2}{\sqrt{2}}+\frac{R_2 +iI_2}{\sqrt{2}},\nn\\
&&\Delta^{0}=\frac{v_3}{\sqrt{2}}+\frac{R_3 +iI_3}{\sqrt{2}}.\nn
\eea
The minimization conditions of eq.~(\ref{eq:2}) are given by,
\bea
&&\mu_1^2=\frac{-2\beta_1 v_1^3-\beta_2 v_1 v_2^2-\beta_3 v_1 v_3^2
  +\kappa v_2^2 v_3}{2v_1},\nn\\
&&\mu_2^2=-\frac{1}{2}\left(2\lambda_1 v_2^2+\beta_2 v_1^2+(\lambda_3+\lambda_5) v_3^2
  -2\kappa v_1 v_3 \right),\\
&&\mu_3^2=\frac{-2(\lambda_2+\lambda_4) v_3^3-(\lambda_3+\lambda_5) v_2^2 v_3
  -\beta_3 v_1^2v_3+\kappa v_1 v_2^2}{2v_3}.\nn
\eea
and from these one can derive a vev seesaw relation of the type
$$ v_1 v_3 \sim  \kappa v_2^2~,$$
where $\kappa$ is the dimensionless coupling that generates the mass
parameter associated to the cubic term in the scalar potential of the
simplest triplet seesaw scheme with explicit \lnv as proposed
in~\cite{Schechter:1980gr} and recently revisited
in~\cite{Bonilla:2015eha}.\\

{\bf Neutral Higgs bosons}\\

One can now write the resulting squared mass matrix for the CP-even
scalars in the weak basis $(R_1, R_2, R_3)$ as follows,
\begin{eqnarray}
\label{eq:4}
  M^2_R =\left[
  \begin{array}{ccc}
   2\beta_1 v_1^2+\frac{1}{2}\kappa v_2^2 \frac{v_3}{v_1} 
 & \beta_2 v_1 v_2 - \kappa v_2 v_3 & \beta_3 v_1 v_3
   -\frac{1}{2}\kappa v_2^2 \\
   \beta_2 v_1 v_2 -\kappa v_2 v_3 & 2 \lambda_1 v_2^2    
 & (\lambda_3+\lambda_5) v_2 v_3 -\kappa v_1 v_2 \\
   \beta_3 v_1 v_3-\frac{1}{2}\kappa v_2^2
 & (\lambda_3+\lambda_5)v_2 v_3-\kappa v_1 v_2 & 2(\lambda_2+
   \lambda_4)v_3^2+\frac{1}{2}\kappa v_2^2\frac{v_1}{v_3}
  \end{array}\right].
  \end{eqnarray}

The matrix $M^2_{R}$ is diagonalized by an orthogonal matrix 
as follows, $\mathcal{O}_{R} M_{R}^2 \mathcal{O}_R^T = 
\text{diag}(m_{H_1}^2,m_{H_2}^2,m_{H_3}^2)$, where
\begin{eqnarray}
 \label{rot1}
  \left( \begin{array}{c} H_1\\ H_2\\ H_3\\ \end{array} \right) = 
 \mathcal{O}_{R} \left( \begin{array}{c} R_1\\ R_2\\ R_3\\ 
 \end{array} \right). 
 \end{eqnarray}

We use the standard parameterization $\mathcal{O}_{R} = 
R_{23} R_{13} R_{12}$ where
\begin{equation}
R_{12} = \left(
\begin{array}{ccc}
c_{12} & s_{12} & 0\\
-s_{12} & c_{12} & 0\\
0 & 0 & 1
\end{array} \right), 
\quad R_{13} = \left(
\begin{array}{ccc}
c_{13} & 0 & s_{13}\\
0 & 1 & 0\\
-s_{13} & 0 & c_{13}
\end{array} \right), 
\quad R_{23} = \left(
\begin{array}{ccc}
1 & 0 & 0\\
0 & c_{23} &  s_{23}\\
0 & -s_{23} & c_{23}
\end{array} \right)
\end{equation}
and $c_{ij} = \cos \alpha_{ij}, s_{ij} = \sin \alpha_{ij}$, so that
the rotation matrix $\mathcal{O}_{R}$ is re-expressed in terms of the
mixing angles in the following way: 
\be
\label{eq:8}
\mathcal{O}_{R} = \left(
\begin{array}{ccc}
c_{12} c_{13}                          &  c_{13} s_{12}                        & s_{13}\\
-c_{23} s_{12} - c_{12} s_{13} s_{23}  & c_{23} c_{12} - s_{12} s_{13}  s_{23}  & c_{13} s_{23}\\
-c_{12} c_{23} s_{13} + s_{23} s_{12}  & -c_{23} s_{12} s_{13} - c_{12} s_{23} & c_{13} c_{23}
\end{array} \right).
\ee

On the other hand, the squared mass matrix for the CP-odd scalars in
the weak basis $(I_1,I_2,I_3)$ is given as,
\begin{eqnarray}
  M^2_I =\kappa\left[
  \begin{array}{ccc}
  \frac{1}{2} v_2^2 \frac{v_3}{v_1} & v_2 v_3      & \frac{1}{2} v_2^2 \\
                v_2 v_3             & 2 v_1 v_3    & v_1 v_2  \\
              \frac{1}{2} v_2^2     &  v_1 v_2      & \frac{1}{2} v_2^2\frac{v_1}{v_3}
  \end{array}\right].
  \end{eqnarray}
  The matrix $M^2_I$ is diagonalized as, $\mathcal{O}_{I} M_{I}^2
  \mathcal{O}_I^T = \text{diag}(0,0,m_A^2)$, where the null masses
  correspond to the would-be Goldstone boson $G^{0}$ and the Majoron
  $J$, while the squared  CP-odd mass is
\be\label{masq} m_{A}^2=\kappa\left(\frac{v_2^2 v_1^2 +v_2^2
      v_3^2 + 4v_3^2 v_1^2}{2v_3 v_1}\right).  \ee 
  The mass eigenstates are linked with the original ones by the
  following rotation,
\begin{eqnarray}\label{rotI}
\left( \begin{array}{c} A_1 \\ A_2 \\ A_3\\ \end{array} \right)\equiv
  \left( \begin{array}{c} J \\ G^{0} \\ A\\ \end{array} \right) = 
 \mathcal{O}_{I} \left( \begin{array}{c} I_1\\ I_2\\ I_3\\ 
 \end{array} \right)
 \end{eqnarray}
where the  matrix $\mathcal{O}_{I}$ is given by,
\begin{eqnarray}
  \label{eq:12}
  O_I =\left[
  \begin{array}{ccc}
  c v_1 V^2  & -2cv_2 v_3^2 & -c v_2^2 v_3 \\
  0          &   v_2/V    & -2v_3/V   \\
  b v_2/2v_1 &     b        & bv_2/2v_3 
  \end{array}\right],
  \end{eqnarray} 
with
\bea
&& V^2=v_2^2+4v_3^2,\nn\\
&& c^{-2}=v_1^2 V^4+4v_2^2 v_3^4+v_2^4 v_3^2\\
&& b^2=\frac{4v_1^2 v_3^2}{v_2^2 v_1^2+v_2^2 v_3^2+4v_3^2 v_1^2}.
\eea
\vskip .5cm
{\bf Charged Higgs bosons}\\

The squared mass matrix for the singly-charged scalar bosons 
in the original weak basis $(\phi^{\pm},\Delta^{\pm})$ is given by,
\begin{eqnarray}
  \label{Mchsq}
  M^2_{H^{\pm}} =\left[
  \begin{array}{cc}
  \kappa v_1 v_3-\frac{1}{2}\lambda_5 v_3^2        & \frac{1}{2\sqrt{2}}v_2(\lambda_5 v_3-2\kappa v_1) \\
  \frac{1}{2\sqrt{2}}v_2(\lambda_5 v_3-2\kappa v_1)& \frac{1}{4v_3}v_2^2(-\lambda_5 v_3+2\kappa v_1)
  \end{array}\right].
  \end{eqnarray}
We now define
\bea
\left(\begin{array}{c}
 G^{\pm} \\
 H^{\pm}
\end{array}
\right)
=\left(\begin{array}{cc}
c_{\pm} & s_{\pm}\\
-s_{\pm} & c_{\pm}
\end{array}
\right)
\left(\begin{array}{c}
\phi^{\pm} \\
\Delta^{\pm}
\end{array}\right),\ \ \text{and}
\quad \mathcal{O}_{\pm} M_{H^{\pm}}^2 \mathcal{O}_\pm^T = 
 \text{diag}(0, m_{H^{\pm}}^2). & 
\eea
where $c_{\pm}$ and $s_{\pm}$ are given as
$c_{\pm}=v_{2}/\sqrt{v_{2}^{2}+2v_{3}^{2}}$ and $
s_{\pm}=\sqrt{2}v_{3}/\sqrt{v_{2}^{2}+2v_{3}^{2}}$. The massless state
corresponds to the would-be Golstone bosons $G^{\pm}$ and the massive state
$H^{\pm}$ is characterized by, 
\be\label{mschsq}
m_{H^{\pm}}^2=\frac{1}{4v_3}(2\kappa v_1-\lambda_5 v_3)(v_2^2+2v_3^2).
\ee 
On the other hand, the doubly-charged scalars $\Delta^{\pm\pm}$ has
mass
\be\label{mdchsq} m_{\Delta^{++}}^2=\frac{1}{2v_3}(\kappa v_1
v_2^2- 2\lambda_4 v_3^3-\lambda_5 v_2^2 v_3).  \ee

\subsection{ Scalar boson mass sum rules}

Notice that using the fact that the smallness of the neutrino mass
implies that the parameters $\kappa$ and $v_3$ are very small one can,
to a good approximation, rewrite eq.~(\ref{eq:4}) schematically in the
form,
\be
M^2_{R}\sim \left(
\begin{array}{ccc}
\star & \star & 0\\
\star & \star & 0\\
   0    & 0 & \star
\end{array} \right)
\ \ \text{so that}\ \
\mathcal{O}_{R}\sim \left(
\begin{array}{ccc}
c_{12}  & s_{12} & 0\\
-s_{12} & c_{12} & 0\\
   0    & 0 & 1
\end{array} \right),
\ee
and eq.~(\ref{masq}) becomes,
\be m_{A}^2\sim \kappa \frac{v_2^2 v_1}{2v_3}. \ee 
As a result, the scalar $H_3$ and the pseudo-scalar $A$ are
almost degenerate,
\be
\label{massrelh3A}
m_{H_3}=(M_R^2)_{33}\approx m_A^2.
\ee
In the same way, by using eqs.~(\ref{masq}), (\ref{mschsq}) 
and (\ref{mdchsq}), one can derive the following mass relations,
\bea
m_A^2-m_{H^+}^2\approx \frac{\lambda_5 v_{2}^2}{4}\ \ \text{and}\ \
2m_{H^+}^2-m_A^2-m_{\Delta^{++}}^2\approx\lambda_{4}v_3^2,\nn
\eea
which can be rewritten in the form,
\be\label{massrel1}
m_{H^+}^2-m_{\Delta^{++}}^2\approx 
m_A^2-m_{H^+}^2\approx \frac{\lambda_5 v_{2}^2}{4}.
\ee
 This sum rule is also satisfied in the Type-II seesaw
  model with explicit breaking of lepton number.  Imposing the
  perturbativity condition one finds that the squared mass difference
  between, say doubly and singly charged scalar bosons, cannot be too
  large~\cite{Bonilla:2015eha}. Explicit comparison shows that
  $\lambda_5$ in eq.~(\ref{eq:2}) corresponds to $\lambda'_{H\Delta}$ in
  Ref.~\cite{Bonilla:2015eha}. Therefore when the couplings of the
  singlet $\sigma$ in eq.~(\ref{eq:2}) are small, $\lambda_5$ is
  constrained to be in the range $[-0.85,0.85]$, so that the remaining
  couplings are kept small up to the Planck scale and vacuum stability
  is guaranteed. See Figure 4 in Ref.~\cite{Bonilla:2015eha}. Likewise
  when one decouples the triplet one also recovers the results found
  in Ref.~\cite{Bonilla:2015kna}.
  
\section{Theoretical constraints}

Before analyzing the sensitivities of the searches for Higgs bosons at
the LHC experiments, we first discuss the restrictions that follow
from the consistency requirements of the Higgs potential.
We can rewrite the dimensionless parameters 
$\lambda_{1,2,3}$ and
$\beta_{1,2,3}$ in eq.~(\ref{eq:2}) in terms of the mixing angles,
$\alpha_{ij}$ and scalar the masses $m_{H_{1,2,3}}$ by solving
$\mathcal{O}_{R} M_{R}^2 \mathcal{O}_R^T =
\text{diag}(m_{H_1}^2,m_{H_2}^2,m_{H_3}^2)$ and $\mathcal{O}_{I}
M_{I}^2 \mathcal{O}_I^T =\text{diag}(0,0,m_A^2)$. Hence one gets,
\begin{small}
\be
\begin{array}{lcl}\label{lambdas1}
\lambda_{1}&=&\frac{1}{2v_{2}^{2}}\left[m_{H_{1}}^{2}c_{13}^{2}s_{12}^{2}
              +m_{H_{2}}^{2}\left(c_{12}c_{23}-s_{12}s_{13}s_{23}\right)^{2}
              +m_{H_{3}}^{2}\left(c_{12}s_{23}+s_{12}s_{13}c_{23}\right)^{2}\right]\nn\\
\lambda_{2}&=&\frac{1}{2v_{3}^{2}}\left[m_{H_{1}}^{2}s_{13}^{2}
              +c_{13}^{2}\left(m_{H_{2}}^{2}s_{23}^{2}+m_{H_{3}}^{2}c_{23}^{2}\right)\right]
              -\left(\lambda_{4}+\kappa \frac{v_{1}v_{2}^{2}}{4v_{3}^{3}}\right)\nn\\
\lambda_{3}&=&\frac{c_{13}}{v_{2}v_{3}}\left[m_{H_{1}}^{2}s_{12}s_{13}
              +m_{H_{2}}^{2}s_{23}\left(c_{12}c_{23}-s_{12}s_{13}s_{23}\right)
              -m_{H_{3}}^{2}c_{23}\left(c_{12}s_{23}+s_{12}s_{13}c_{23}\right)\right]\nn\\
              &&-\left(\lambda_{5}-\kappa \frac{v_{1}}{v_{3}}\right)\\
\beta_{1}&=&\frac{1}{2v_{1}^{2}}\left[m_{H_{1}}^{2}c_{12}^{2}c_{13}^{2}
              +m_{H_{2}}^{2}\left(s_{12}c_{23}+c_{12}s_{23}s_{13}\right)^{2}
              +m_{H_{3}}^{2}\left(s_{12}s_{23}-c_{12}c_{23}s_{13}\right)^{2}\right]
              -\kappa\frac{v_{2}^{2}v_{3}}{4v_{1}^{3}}\nn\\
\beta_{2}&=&\frac{1}{v_{1}v_{2}}\left[m_{H_{1}}^{2}c_{12}s_{12}c_{13}^{2}
              -m_{H_{2}}^{2}(c_{23}s_{12}+c_{12}s_{13}s_{23})(c_{12}c_{23}
              -s_{12}s_{13}s_{23})\right.\nn\\\
              & &\left.-m_{H_{3}}^{2}(s_{23}s_{12}-c_{12}s_{13}c_{23})(c_{12}s_{23}
              +s_{12}s_{13}c_{23})\right]+\kappa\frac{v_{3}}{v_{1}}\nn\\
\beta_{3}&=&\frac{c_{13}}{v_{1}v_{3}}\left[m_{H_{1}}^{2}c_{12}s_{13}
              -m_{H_{2}}^{2}s_{23}(s_{12}c_{23}+c_{12}s_{13}s_{23})
              +m_{H_{3}}^{2}c_{23}(s_{12}s_{23}-c_{12}s_{13}c_{23})\right]
              +\kappa\frac{v_{2}^{2}}{2v_{1}v_{3}}.\nn
\end{array}
\ee
\end{small}

In addition, using eqs.~(\ref{masq}), (\ref{mschsq}) and (\ref{mdchsq}) we 
can write the dimensionless parameters $\lambda_{4,5}$ and $\kappa$ as 
functions of the vevs $v_{1,2,3}$ and the masses of the pseudo-,
singly- and doubly-charged scalar bosons (i.e. $m_{A}$, $m_{H^{\pm}}$ and 
$m_{\Delta^{\pm\pm}}$, respectively) as, 
\bea\label{lambdas2}
\lambda_{4}&=&\frac{1}{v_{3}^{2}}\left(2 m_{H^{\pm}}^{2}\frac{v_{2}^{2}}{v_{2}^{2}
            +2v_{3}^{2}}-m_{A}^{2}\frac{v_{1}^{2}v_{2}^{2}}{v_{2}^{2}v_{3}^{2}
            +v_{1}^{2}(v_{2}^{2}+4v_{3}^{2})}-m_{\Delta^{\pm\pm}}^{2}\right)\nn\\
\lambda_{5}&=&\left(-4m_{H^{\pm}}^{2}\frac{1}{v_{2}^{2}+2v_{3}^{2}}
            +4m_{A}^{2}\frac{v_{1}^{2}}{v_{2}^{2}v_{3}^{2}+v_{1}^{2}(v_{2}^{2}
            +4v_{3}^{2})}\right)\label{l4l5k}\\
\kappa     &=& 2m_{A}^{2}\frac{v_{1}v_{3}}{v_{2}^{2}v_{3}^{2}
            +v_{1}^{2}(v_{2}^{2}+4v_{3}^{2})}.\nn
\eea

From the theoretical side we have to ensure that the scalar potential in 
the model is bounded from below (BFB). 

\subsection{Boundedness Conditions}

In order to ensure that the scalar potential in eq.~(\ref{eq:2}) 
is bounded from below we have to derive the conditions on the 
dimensionless parameters such the quartic part of the scalar
potential is positive $V^{(4)}>0$ as the fields go to infinity.
We have that the parameter $\kappa\ll1$ (due to the smallness 
of the neutrino mass) and non-negative. This follows from
\be\label{kapma}
\kappa\approx2m_A^2 \frac{v_3}{v_1 v_2^2}.
\ee
where we have used the last expression in eq.~(\ref{lambdas2}) and the
fact that $v_3\ll v_2,v_1$. Then $\kappa$ is neglected with respect to
the other dimensionless parameters $\lambda_i$ and $\beta_j$,
i.e. $\lambda_i,\,\beta_j\gg\kappa$. As a result the quartic part of
the potential $V^{(4)}|_{\kappa=0}$ turns to be a biquadratic form
$\lambda_{ij}\varphi_i^2\varphi_j^2$ of real fields.  Therefore, in
this strict limit, the copositivity criteria described
in~\cite{Kannike:2012pe} may be applied and the boundedness conditions
for eq.~(\ref{eq:2}) are the following, \bea\label{bfbconds} & &
\lambda_{1}>0,\ \ \beta_{1}>0, \ \ \lambda_{24}>0,\ \
\hat{\lambda}\equiv\beta_{2}+2\sqrt{\beta_{1}\lambda_{1}}>0,\nn\\
& & \tilde{\lambda}\equiv\beta_{3}+2\sqrt{\beta_{1}\lambda_{24}}>0,\ \
\bar{\lambda}\equiv\lambda_{3}+\theta(-\lambda_5)\lambda_5+
2\sqrt{\lambda_{1}\lambda_{24}}>0, \ \ \text{and}\ \ \nn\\
& & \sqrt{\beta_1 \lambda_1
  \lambda_{24}}+\left[\lambda_3+\theta(-\lambda_5)
  \lambda_5\right]\sqrt{\beta_1}+ \beta_2 \sqrt{\lambda_{24}} +\beta_3
\sqrt{\lambda_1} +\sqrt{\hat{\lambda}\tilde{\lambda}\bar{\lambda}}>0,
\eea where $\lambda_{24}\equiv \lambda_2+\lambda_4$.
In addition all the dimensionless parameters in the scalar potential
are required to be less than $\sqrt{4\pi}$ in order to fulfill the
perturbativity condition.
 
\subsection{Astrophysical constraints}

In our type-II seesaw model there are some constraints on the
magnitude of $SU(2)$ triplet's vev $\vev{ \Delta}=v_3$, that one must
take into account.
First of all, $v_{3}$ is constrained to be smaller than a few GeVs due
to the $\rho$ parameter ( $\rho=1.0004\pm0.00024$
~\cite{Agashe:2014kda}).

On the other hand, the presence of the Nambu-Goldstone boson
associated to spontaneous \lnv and neutrino mass generation implies
that there is a most stringent constraint on $v_{3}$ coming from
astrophysics, due to supernova cooling. If the majoron is a strict
Goldstone boson (or lighter than typical stellar temperatures) one has
an upper bound for the Majoron-electron coupling
$$|g_{Jee}| \lsim 10^{-13},$$
This is discussed, for example, in Ref.~\cite{Choi:1989hi} and
references therein.  This implies
$$|g_{Jee}|=|\mathcal{O}^{I}_{12}m_{e}/v_{2}|.$$ Taking into account the profile 
of the Majoron~\cite{Schechter:1981cv}~\footnote{This is derived
  either by explicit analysis of the scalar potential or simply by
  symmetry, using Noether's theorem~\cite{Schechter:1981cv}.} one can
translate this as a bound on the projection of the Majoron onto the
doublet as follows~\cite{Diaz:1998zg}
\begin{equation}
|\vev{ J|\phi}|=\frac{2|v_2|v_3^2}{\sqrt{v_1^2 
(v_2^2+4v_3^2)^2+4 v_2^2 v_3^4+v_2^4 v_3^2}}\lesssim10^{-7}.  
\end{equation}

\begin{figure}[H]
\vspace{-2pt}
  \centering
 \includegraphics[width=0.4\textwidth]{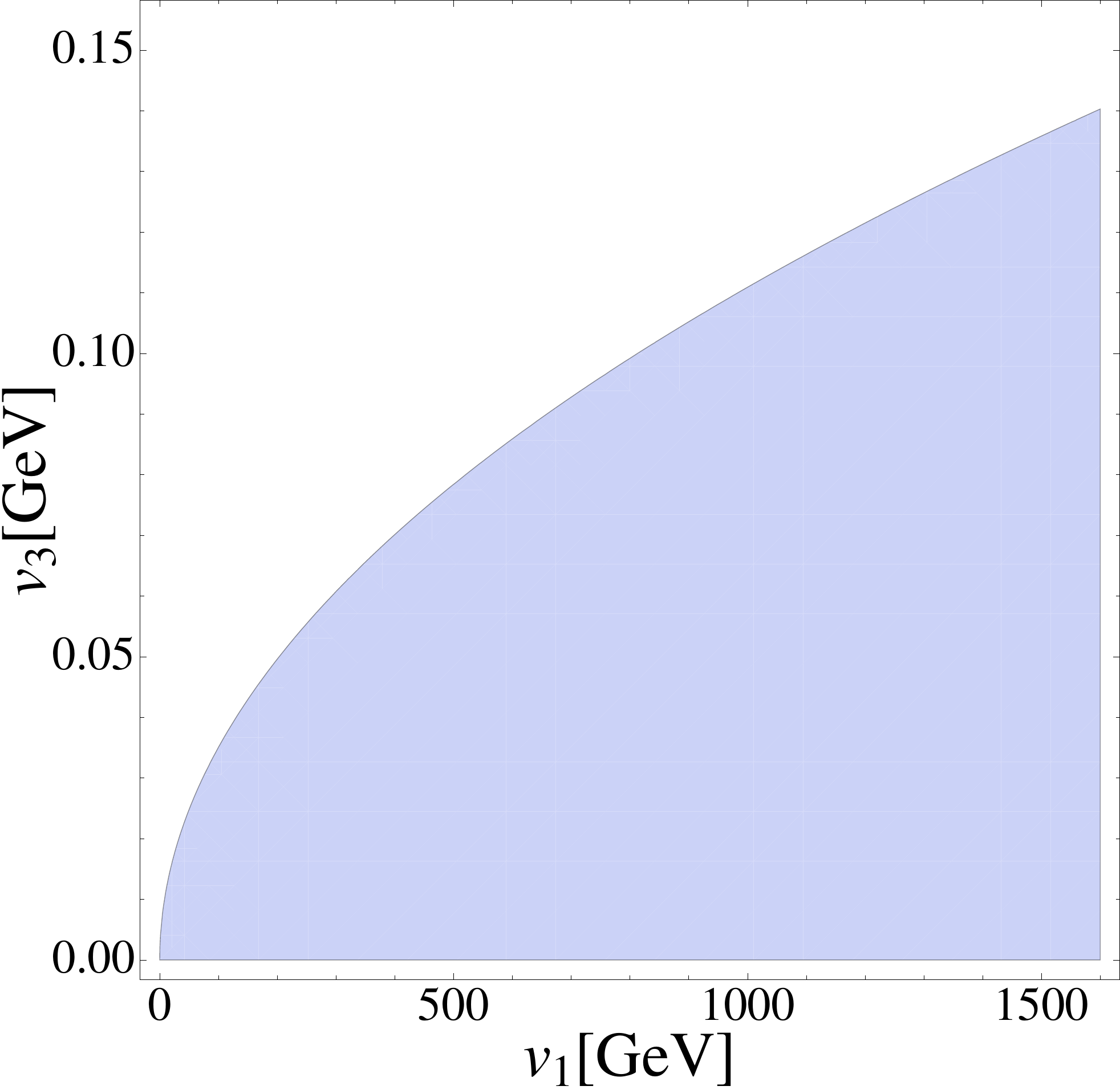}
\vspace{-5pt}
\caption{The shaded region represents the allowed region of $v_3$ as
  function of $v_1$. \label{Fv1v3}}
\end{figure}

Notice that this restriction on the triplet's vev is stronger that the
one stemming from the $\rho$ parameter.  The shaded region in
Fig.~\ref{Fv1v3} corresponds to the allowed region of $v_{3}$ as
function of $v_{1}$.
To close this section we mention that our phenomenological analysis
remains valid if the Nambu-Goldstone boson picks up a small mass from,
say, quantum gravity effects.

\section{Type-II seesaw Higgs searches at the LHC}
\label{sec:exper-sens}

We now turn to the study of the experimental sensitivities of the LHC
experiments to the parameters characterizing the ``123'' type-II
majoron seesaw Higgs sector, as proposed in~\cite{Schechter:1981cv}.
In the following we will assume that $m_{H_1}<m_{H_2}<m_{H_3}$ where
1,2,3 refer to the mass ordering in the CP even Higgs sector.
Therefore, there are two possible cases that can be
considered\footnote{Recall that $m_{H_{3}} \approx m_A$,
  eq.~(\ref{massrelh3A}), which implies that the mass of $H_3$ must be
  close to that of the doubly-charged scalar mass. Therefore, as we
  will see in the next section, the existing bounds on searches of the
  doubly-charged scalar exclude the case where $m_{H_{3}}$ is lighter
  than the other CP-even mass eigenstates.}:
\begin{itemize}
 \item[{\bf (i)}] $m_{H_1}<m_H$ and $m_{H_2}=m_H$;
 \item[{\bf (ii)}] $m_{H_1}=m_H$,
\end{itemize}
where $m_H$ is the mass of the Higgs reported by the
ATLAS~\cite{Aad:2014eva} and CMS~\cite{Khachatryan:2014ira}
collaborations, i.e. $m_{H}=125.09\pm0.21(\text{stat.})
\pm0.11(\text{syst.})~\g$~\cite{Aad:2015zhl}.  For case (i), we have
to enforce the constraints coming from LEP-II data on the lightest
CP-even scalar coupling to the SM and those coming from the LHC Run-1
on the heavier scalars.  Such situation has been discussed by us in
Ref.~\cite{Bonilla:2015uwa} in the simplest ``12-type'' seesaw Majoron
model. In case (ii), only the constraints coming from the LHC must be
taken into account.

The neutral component of the \sm Higgs doublet couplings get modified
as follows, 
\be\label{htosm} 
\phi^0 \to C_1 H_1 +C_2 H_2 +C_3 H_3 \ee
where we have defined $C_i\equiv\mathcal{O}^R_{i2}$ and 
$\mathcal{O}^R_{ij}$ are the matrix elements of $\mathcal{O}^R$ 
in eq.~(\ref{eq:8}).

\subsection{LEP constraints on invisible Higgs decays}
\label{sec:lep-constraints}

The constraints on $H_1$, when $m_{H_1}<125\g$, stem from the 
process $e^+ e^- \to Z h \to Z b\bar{b}$ which is written 
as~\cite{Abdallah:2004wy}
\begin{align}
\label{eq:defsigma}
\sigma_{hZ \to b\bar{b}Z} = &\sigma_{h Z}^{SM}\times R_{h Z}\times
BR(h \to b \bar{b}) \nn\\ 
= &\sigma_{hZ}^{SM}\times C^2_{Z(h \to  b\bar{b}) }\,, 
\end{align}
where $\sigma_{hZ}^{SM}$ is the SM $hZ$ cross section, $R_{hZ}$ is the
suppression factor related to the coupling of the Higgs
boson\footnote{The Feynman rules for the couplings of the Higgs bosons
  $H_{i}$ to the $Z$ are the following:
  $i\frac{g^2}{2c_W^2}\left(\mathcal{O}^R_{i2}
    v_2+\mathcal{O}^R_{i3}v_3\right)g_{\mu\nu}$} to the gauge boson
$Z$. Since $v_3\ll v_2$, we have that the factor $R_{H_{i}Z}\approx
C_i^2$ where $C_1=\cos\alpha_{13} \sin\alpha_{12}$,
eq.~(\ref{htosm}). Notice that $C_1\approx \sin\alpha_{12}$ for
the limit $\alpha_{13}\ll1$ and then one obtains the same exclusion 
region depicted in Fig.~1 in Ref.~\cite{Bonilla:2015uwa}.

\subsection{LHC constraints on the Higgs signal strengths}
\label{sec:lhc-constraints-on}

In addition, we have to enforce the limits coming from the \sm decay
channels of the Higgs boson. These are given in terms of the signal
strength parameters,
\begin{equation}
  \label{eq:16}
  \mu_f =
\frac{\sigma^\textrm{NP}(pp \to h)}{\sigma^\textrm{SM}(pp \to h)}\,
\frac{BR^\textrm{NP}(h \to f)}{BR^\textrm{SM}(h \to f)} ,
\end{equation}
where $\sigma$ is the cross section for Higgs production, $BR(h \to
f)$ is the branching ratio into the \sm final state $f$, the labels NP
and SM stand for New Physics and Standard Model respectively.  These
can be compared with those given by the experimental
collaborations. The most recent results of the signal strengths from a
combined ATLAS and CMS analysis~\cite{ATLAS-CONF-2015-044} are shown
in Table~\ref{tab:1}.
\begin{table}[H]
\centering
\begin{tabular}{|ccccccc|}
\hline
channel & & ATLAS  & & CMS & & ATLAS+CMS \\
\hline
$\mu_{\gamma\gamma}$  & &
$1.15^{+0.27}_{-0.25}$  & &
$1.12 ^{+0.25}_{-0.23}$ & &
$1.16^{+0.20}_{-0.18}$
\\*[2mm]
$\mu_{WW}$  & &
$1.23^{+0.23}_{-0.21}$ & &
$0.91^{+0.24}_{-0.21}$  & &
$1.11^{+0.18}_{-0.17}$
\\*[2mm]
$\mu_{ZZ}$  & &
$1.51^{+0.39}_{-0.34}$ & &
$1.05^{+0.32}_{-0.27}$ & &
$1.31^{+0.27}_{-0.24}$
\\*[2mm]
$\mu_{\tau\tau}$  & &
$1.41^{+0.40}_{-0.35}$ & &
$0.89^{+0.31}_{-0.28}$ & &
$1.12^{+0.25}_{-0.23}$
\\*[2mm]
$\mu_{b b}$  & &
$0.62^{+0.37}_{-0.36}$ & &
$0.81^{+0.45}_{-0.42}$ & &
$0.69^{+0.29}_{-0.27}$
\\*[2mm]
\hline
\end{tabular}
\caption{\label{tab:1} Current experimental results  of
  ATLAS and CMS, Ref.~\cite{ATLAS-CONF-2015-044}.}
\end{table}
One can see with ease that the LHC results indicate that $\mu_{VV}
\sim 1$. In our analysis, we assume that the LHC allows deviations up
to 20\% as follows,
\begin{equation}
  \label{eq:17}
  0.8 \leq \mu_{XX} \leq 1.2
\end{equation}

\subsection{LHC bounds on the heavy neutral scalars}
\label{sec:bounds-heavy-neutral}

In our study we will impose the constraints on the heavy scalars from
the recent LHC scalar boson searches. Therefore, we use the bounds set
by the search for a heavy Higgs in the $H\to WW$ and $H\to ZZ$ decay
channels in the range $[145-1000]\g$ \cite{Khachatryan:2015cwa} and in
the $h\to \tau\tau$ decay channel in the range range $[100-1000]\g$
\cite{Khachatryan:2014wca}.  We also adopt the constraints on the
process $h\to \gamma \gamma$ in the range $[65 -600]\g$
\cite{Aad:2014ioa} and the range $[150,850]\g$
\cite{Khachatryan:2015qba}. Besides, we impose the bounds in the $A
\to Z h$ decay channel in the range $[220-1000]\g$ \cite{Aad:2015wra}.

\subsection{Summary of the searches of charged scalars}
\label{chargedhiggs}
 
The type-II seesaw model with explicit breaking of lepton number
contains seven physical scalars: two CP-even neutral scalars $H_1$ and
$H_2$, one CP-odd scalar $A$ and four charged scalars $\ddpm$ and
$\hpm$. Such a scenario has been widely studied in the literature and
turns out to be quite appealing because it could be tested at the LHC
\cite{Garayoa:2007fw, Perez:2008zc,Perez:2008ha,
  delAguila:2008cj,Aoki:2011pz, Akeroyd:2012ms,Dev:2013ff,
  delAguila:2013mia,Chen:2014qda,Kanemura:2014goa, Han:2015hba}.
 For instance, the existence of charged scalar
bosons provides additional contributions to the one-loop decays of the
\sm Higgs boson. Indeed, they could affect the one-loop decays
$h\to\gamma\gamma$ \cite{Akeroyd:2012ms,Dev:2013ff} and $h\to Z\gamma$
\cite{Dev:2013ff} in a substantial way.  In this case the signal
strength $\mu_{\gaga}$ can set bounds on the
mass of the charged scalars, $\Delta^{\pm\pm}$ and/or $H^\pm$.\\

The doubly-charged scalar boson has the following possible decay
channels: $\ell^{\pm}\ell^\pm$, $W^\pm W^\pm$, $W^\pm H^\pm$ and
$H^\pm H^\pm$. However, it is known that for an approximately
degenerate triplet mass spectrum and vev $v_3\lsim 10^{-4}\g$ the
doubly charged Higgs coupling to $W^\pm$ is suppressed (because it is
proportional to $v_3$ as can be seen from Table~\ref{tab:2}) and hence
$\ddpm$ predominantly decays into like-sign dileptons~
\cite{Sugiyama:2012yw,delAguila:2013mia,Han:2015hba}. In this case,
CMS \cite{Chatrchyan:2012ya} and ATLAS \cite{ATLAS:2012hi} have
currently excluded at $95\%$~C.L., depending on the assumptions on the
branching ratios into like-sign dileptons, doubly-charged masses
between 200 and 460~$\g$~\footnote{From doubly-charged scalar boson
  searches performed by ATLAS and CMS one can also constrain the
  lepton number violation processes
  $pp\to\ddpm\Delta^{\mp\mp}\to\ell^{\pm}\ell^{\pm}W^{\mp}W^{\mp}$ and
  $pp\to\ddpm H^{\mp}\to\ell^{\pm}\ell^{\pm}W^{\mp}Z$
  \cite{delAguila:2013mia}.  This may also shed light on the Majorana
  phases of the lepton mixing
  matrix~\cite{Garayoa:2007fw,Perez:2008zc,Perez:2008ha}.}.  For
$v_3\gsim 10^{-4}\g$, the Yukawa couplings of triplet to leptons are
too small so that $\Delta^{\pm\pm}$ dominantly decays to like-sign
dibosons, in which case the collider limits are rather
weak~\cite{Kanemura:2013vxa,Kanemura:2014ipa,
  Kanemura:2014goa,Khachatryan:2014sta}.\\

In the present ``123'' type-II seesaw model there are two additional
physical scalars, a massive CP-even scalar $H_3$ and the massless
majoron $J$.  The latter, associated to the spontaneous breaking
of lepton number, provides non-standard decay channels of other Higgs
bosons as missing energy in the final state \footnote{
  These include, for example, $H_i\to JJ$ and $H^\pm\to J
  W^{\mp}$. Here we focus mainly on the first, the decays of $H^\pm$
  deserve further study but it is beyond the scope of this
  work and will be considered elsewhere.
}. 

\section{Invisible Higgs decays at the LHC}
\label{sec:non-standard-model}

We now turn to the case of genuinely non-standard Higgs decays.  We
focus on investigating the LHC sensitivities on the invisible Higgs
decays. In so doing we take into account how they are constrained by
the available experimental data.  In the previous section we mentioned
that in our study the CP- even scalars obey the following mass
hierarchy $m_{H_{1}}< m_{H_{2}}<m_{H_{3}}$. Furthermore, we will also
assume that the masses $m_{H_{3}}$, $m_{A}$, $m_{H^{+}}$ and
$m_{\Delta^{++}}$ are nearly degenerate.
As a consequence, the decay of any CP-even Higgs $H_i$ into the
pseudo-scalar $A$ is not kinematically allowed.  Therefore, the new
decay channels of the CP-even scalars are just, $H_{i}\to JJ$ and
$H_{i}\to 2H_{j}$ (when $m_{H_{i}}< \frac{m_{H_{j}}}{2}$ for $i\neq
j$). The latter contributing also to the invisible decay channel of
the Higgs as, $H_{i}\to 2H_{j}\to 4J$.

The Higgs-Majoron couplings are given by,
\be\label{ghJJ}
g_{H_{a}JJ}=\left(\frac{(\mathcal{O}^{I}_{12})^2}{v_{2}}\mathcal{O}^{R}_{a2}+
            \frac{(\mathcal{O}^{I}_{13})^2}{v_{3}}\mathcal{O}^{R}_{a3}+
            \frac{(\mathcal{O}^{I}_{11})^2}{v_{1}}\mathcal{O}^{R}_{a1}
            \right)m_{H_{a}}^2,
\ee
where $\mathcal{O}^{I}_{ij}$ are the elements of the rotation 
matrix in eq.~(\ref{eq:12}) and the decay width is given by
\begin{equation}
  \label{eq:13}
\Gamma(H_a \to JJ) = \frac{1}{32\pi} \frac{g^2_{H_{a} JJ}}{m_{H_a}}
\, .
\end{equation}
{ 
Following our conventions we have that the trilinear coupling 
$H_{2}H_{1}H_{1}$ turns out to be,
\begin{small}
\bea
\frac{g_{H_{2}H_{1}H_{1}}}{2}&&= 3\lambda_{1}(\mathcal{O}^{R}_{12})^2 \mathcal{O}^{R}_{22} v_{2}+
     3(\lambda_{2}+\lambda_{4})(\mathcal{O}^{R}_{13})^2 \mathcal{O}^{R}_{23}v_{3}\nn\\
&&+ \frac{(\lambda_{3}+\lambda_{5})}{2}\left[(\mathcal{O}^{R}_{13})^2 \mathcal{O}^{R}_{22} 
     v_{2}+(\mathcal{O}^{R}_{12})^2 \mathcal{O}^{R}_{23} v_{3}+2 \mathcal{O}^{R}_{12} 
     \mathcal{O}^{R}_{13} (\mathcal{O}^{R}_{23} v_{2}+\mathcal{O}^{R}_{22} v_{3})\right]\nn\\
&&+   3\beta_{1}(\mathcal{O}^{R}_{11})^2 \mathcal{O}^{R}_{21} v_{1}+
      \frac{\beta_{2}}{2}\left[(\mathcal{O}^{R}_{12})^2 \mathcal{O}^{R}_{21} v_{1}
      +(\mathcal{O}^{R}_{11})^2 \mathcal{O}^{R}_{22} v_{2}+
      2 \mathcal{O}^{R}_{11} \mathcal{O}^{R}_{12} (\mathcal{O}^{R}_{22} v_{1}
      +\mathcal{O}^{R}_{21} v_{2})\right]\nn\\
&&+   \frac{\beta_{3}}{2}\left[(\mathcal{O}^{R}_{13})^2 \mathcal{O}^{R}_{21} v_{1}
      +(\mathcal{O}^{R}_{11})^2 \mathcal{O}^{R}_{23} v_{3}+
      2\mathcal{O}^{R}_{11} \mathcal{O}^{R}_{13} (\mathcal{O}^{R}_{23} v_{1}+
      \mathcal{O}^{R}_{21} v_{3}))\right]\nn\\
&&+   \frac{\kappa}{2}\left[- 2\mathcal{O}^{R}_{11}\mathcal{O}^{R}_{13}
      \mathcal{O}^{R}_{22}v_{2}-(\mathcal{O}^{R}_{12})^2
      (\mathcal{O}^{R}_{23} v_{1}+\mathcal{O}^{R}_{21} v_{3})
     -2 \mathcal{O}^{R}_{12} (\mathcal{O}^{R}_{13} (\mathcal{O}^{R}_{22} v_{1}
     +\mathcal{O}^{R}_{21} v_{2})\right.\nn\\
&&+   \left. \mathcal{O}^{R}_{11} (\mathcal{O}^{R}_{23} 
      v_{2}+\mathcal{O}^{R}_{22} v_{3}))\right].
\eea
\end{small}
and hence, for example when $m_{H_{1}}<2m_{H_2}$, the decay width 
$H_{2}\to H_{1}H_{1}$ is given by 
\be
\Gamma\left(H_{2}\to H_{1}H_{1}\right)=\frac{g_{H_{2}H_{1}H_{1}}^{2}}
{32 \pi m_{H_{2}}}\left(1-\frac{4m_{H_{1}}^{2}}{m_{H_{2}}^{2}}\right)^{1/2}.
\ee

As we already mentioned, a salient feature of adding an isotriplet to
the \sm is that some $visible$ decay channels of the Higgs receive
further contributions from the charged scalars, namely the one-loop
decays $h\to\gamma\gamma$ and $h\to Z\gamma$.  That is, the scalars
$H^{\pm}$ and $\Delta^{\pm\pm}$ contribute to the one-loop coupling of
the Higgs to two-photons and to $Z$-photon, leading to deviations from
the \sm expectations for these decay channels. The interactions
between CP-even and charged scalars are described by the following
vertices,  
\bea
& H_{a}H^{+}H^{-}& :\quad i g_{H_{a}H^{+}H^{-}}\notag\\
& H_{a}\Delta^{++}\Delta^{--}& : \quad i
g_{H_{a}\Delta^{++}\Delta^{--}} \notag \eea 
where
\begin{small}
\bea
g_{H_{a}H^{+}H^{-}}&=&\frac{1}{2(v_{2}^{2}+2v_{3}^{2})}\left[8\lambda_{1}
\mathcal{O}^{R}_{a2}v_{2}v_{3}^{2}+4(\lambda_{2}+\lambda_{4})\mathcal{O}^{R}_{a3}v_{2}^{2}v_{3}
+2\lambda_{3}(\mathcal{O}^{R}_{a2}v_{2}^{3}+2 \mathcal{O}^{R}_{a3}v_{3}^{3})\right.\nn\\
&+&\lambda_{5}v_{2}[-2\mathcal{O}^{R}_{a3}v_{2}v_{3}+\mathcal{O}^{R}_{a2}(v_{2}^{2}-2v_{3}^{2})]
+4\beta_{2}\mathcal{O}^{R}_{a1}v_{1}v_{3}^{2}+2\beta_{3}\mathcal{O}^{R}_{a1}v_{1}v_{2}^{2}\nn\\
&+&\left.4\kappa v_{2}v_{3}(\mathcal{O}^{R}_{a2}v_{1}+\mathcal{O}^{R}_{a1}v_{2})\right]\nn\\
g_{H_{a}\Delta^{++}\Delta^{--}}&=&2\lambda_{2}\mathcal{O}^{R}_{a3}v_{3}+\lambda_{3}
\mathcal{O}^{R}_{a2}v_{2}+\beta_{3}\mathcal{O}^{R}_{a1}v_{1}.\nn
\eea
\end{small}

Note that the contributions of $\hpm$ and $\ddpm$ to the decays
$h\to\gaga$ and $h\to Z\gamma$ are functions of the singlet's vev
$v_1$, this is in contrast to what happens in the type-II seesaw model
with explicit violation of lepton number.  According to
eq.~(\ref{bfbconds}) the dimensionless parameters $\lambda_i$ and
$\beta_i$ can change the sign of the couplings of
$g_{H_{a}H^{+}H^{-}}$ and $g_{H_{a}\Delta^{++}\Delta^{--}}$, hence the
contribution of the charged scalars to $h\to\gaga$ and $h\to Z\gamma$
may be either constructive or destructive.

For the computation of the decay widths $h\to \gamma\gamma$ and $h\to
Z\gamma$ we use the expressions and conventions given in
Ref.~\cite{Fontes:2014xva}. The decay width $\Gamma(H_a\to
\gamma\gamma)$ turns out to be
\begin{equation}
 \Gamma(H_a\to \gamma\gamma)= \frac{G_F\alpha^2 m_{H_a}^3}{128\sqrt{2}\pi^3}
  \left|X_F^{\gamma\gamma}+X_W^{\gamma\gamma}+X_{H}^{\gamma\gamma}\right|^2
\end{equation}
where $G_F$ is the Fermi constant, $\alpha$ is the
fine structure constant and the form factors $X_{i}^j$ are given by
\footnote{
We have taken into account that $v_{3}$ is very small so that
any contribution involving the triplet's vev is neglected. Then
for instance the Feynman rule for the vertex $H_a W^+_\mu W^-_\nu\,\, 
:\,\,i\frac{g^{2}}{2}(O^{R}_{a2} v_{2}+2 O^{R}_{a3} v_{3})\gmn$, 
is approximated as $\sim i\frac{g^{2}}{2}(O^{R}_{a2} v_{2})\gmn$
(see Table~\ref{tab:2}).},
\begin{eqnarray}\label{ffactor1}
X_F^{\gaga}&=& -2 C_a\sum_f  N_c^f Q_f^2 \tau_f \left[1+(1-\tau_f)f(\tau_f)\right],
\nn\\
X_{W}^{\gaga}&=& C_a \left[2+\tau_W+3\tau_W(2-\tau_W)f(\tau_W)\right]
\\
X_{H}^{\gaga}&=&
            -\frac{g_{H_{a}H^{+}H^{-}} v}{2m^2_{H^{\pm}}}\tau_{H^{\pm}}\left[
            1-\tau_{H^{\pm}}f(\tau_{H^{\pm}})\right]
          -4\frac{g_{H_{a}\Delta^{++}\Delta^{--}} v}{2m^2_{\Delta^{\pm\pm}}}\tau_{\Delta^{\pm\pm}}\left[
          1-\tau_{\Delta^{\pm\pm}}f(\tau_{\Delta^{\pm\pm}})\right].
\nn
\end{eqnarray}
where $\tau_x=4m_x^2/m_{Z}^2$. Here $N_c^F$ and $Q_F$ denote,
respectively, the number of colors and electric charge of a given
fermion.  The one-loop function $f(\tau)$ is defined in appendix~
\ref{loopfuncs}.  The parameters $C_a$ correspond to the \sm Higgs
couplings in eq.~(\ref{htosm}). \\

The decay width $\Gamma(H_a\to \zg)$, using the notation in
Ref.~\cite{Fontes:2014xva}, is expressed as follows
\begin{equation}
 \Gamma(H_a\to \zg)= \frac{G_F\alpha^2 m_{H_a}^3}{64\sqrt{2}\pi^3}
  \left(1-\frac{m_Z^2}{m_{H_{a}}^2}\right)^3
  \left|X_F^{\zg}+X_W^{\zg}+X_{H}^{\zg}\right|^2
\end{equation}
where the form factors $X_i^j$ are given by\footnote{ Here we have
  also assumed $v_3\ll1$ so as to make the following approximation,
  $H^{+}H^{-}Z_\mu:-i g\sin\theta_W \tan\theta_W (p_+ -p_-)_\mu$.}  ,
\be
\begin{array}{lcl}\label{ffactor2}
X_F^{\zg}&=& - 4C_a\sum_f  N_c^f \frac{g_V^f Q_f m_f^2}{s_W c_W} 
    \left\{\frac{2m_Z^2}{(m_{H_a}^2-m_Z^2)^2}\Delta B_0^f +
    \frac{1}{m_{H_a}^2-m_Z^2}\right.
    \nn\\
    &\times&\left.\left[ (4m_f^2-m_{H_a}^2+m_Z^2)C_0^f+2
    \right]\right\}
\nn\\
X_{W}^{\zg}&=&\frac{C_a}{\tan\theta_W} \left\{\frac{1}{(m_{H_a}^2-m_Z^2)^2}
\left[m_{H_a}^2(1-\tan^2\theta_W)-2m_W^2(-5+\tan^2\theta_W)\right]m_Z^2 \Delta B_0^W\right.
\nn\\
&+& \frac{1}{(m_{H_a}^2-m_Z^2)}\left[m_{H_a}^2(1-\tan^2\theta_W)-2m_W^2(-5+\tan^2\theta_W)
\right.
\\
&+&2m_W^2\left.\left[ (-5+\tan^2\theta_W)(m_{H_a}^2-2m_W^2)-2m_Z^2(-3+\tan^2\theta_W)\right]C_0^W
\right]
\nn\\
X_{H}^{\zg}&=&
            -2g_{H_{a}H^+ H^-} v \frac{\tan\theta_W}{(m_{H_a}^2-m_Z^2)}
            \left[\frac{m_Z^2}{m_{H_a}^2-m_Z^2}\Delta B_0^\pm + (2 m_{H^{\pm}}^2 C^\pm_0+1) \right]
\nn\\
           &-&4\frac{g_{H_{a}\Delta^{++}\Delta^{--}} v}{\tan\theta_W}
	   \frac{(1-\tan^2\theta_W)}{(m_{H_a}^2-m_Z^2)}
            \left[\frac{m_Z^2}{m_{H_a}^2-m_Z^2}\Delta B_0^{\pm\pm} + (2 m_{\Delta^{\pm\pm}}^2 
	    C^{\pm\pm}_0+1) \right]
\nn	    
\end{array}
\ee
where $C^{b}_0$ and $\Delta B^b_{0}$ are defined in 
appendix~\ref{loopfuncs}.
}
\section{Type-II seesaw neutral Higgs searches at the LHC}

We stated above that in our study we are assuming $m_{H_{1}}
<m_{H_{2}}<m_{H_{3}}$ and $v_1\gtrsim v_2$. Furthermore, because of
the $\rho$ parameter and the astrophysical constraint on the triplet's
vev we also have that $v_3\ll v_1,v_2$.  We found that the smallness
of $v_3$ and the perturbativity condition of the potential lead to a
very small mixing between the mass eigenstate $H_{3}$ and the CP-even
components of the fields, $\sigma$ and $\Phi$, in other words, the
angles $\alpha_{13}$ and $\alpha_{23}$ must lie close to 0 or
$\pi$. As a result, we obtain the following relation,
\be\label{massrel2}
m_{H_{3}}^{2}-m_A^2\simeq 2\lambda_2 v_3^2\Longrightarrow m_{H_3}\simeq m_{A}.
\ee
This extra mass relation is derived from eq.~(\ref{lambdas1}), by using
eq.~(\ref{kapma}) and the fact that $\alpha_{13,23} \sim 0(\pi)$.
In addition, also as a result of $\alpha_{13,23}\sim 0(\pi)$, we find
that the coupling of $H_3$ to the \sm states is negligible,
\begin{equation}
 \frac{g_{H_{3}ff}}{g^{\text{SM}}_{hff}}=
 \frac{g_{H_{3}VV}}{g^{\text{SM}}_{hVV}}=C_3\sim0.
\end{equation}

In Fig.~\ref{fig:ms} of appendix~\ref{massspec} we give a schematic
illustration of the mass profile of the Higgs bosons in our model. The
mass spectrum and composition are summarized in Table~\ref{tab:ms}, and
provide a useful picture in our following analyses.

\subsection{Analysis (i)}
\label{sec:analysis-i}

In this case we have taken the isotriplet vev $v_{3}=10^{-5}~\g$,
automatically safe from the constraints stemming from astrophysics and
the $\rho$ parameter. We have also considered the following mass
spectrum, 
\bea &m_{H_{1}}=[15,115]~\g,\ \ m_{H_{2}}=125~\g,\ \
m_{H_{3}}\simeq m_{A}\simeq m_{H^{\pm}}\simeq
m_{\Delta^{\pm\pm}}=500\g,\nn \eea 
and varied the parameters as
\bea
&v_{1}\in[100,2500]~\g,\ \ \alpha_{12}\in[0,\pi]\ \ \text{and}\ \
\alpha_{13,23}= \delta_\alpha\,(\pi-\delta_\alpha) \eea 
where $0\leq\delta_\alpha<0.1$. As described in
section~\ref{sec:exper-sens} we must enforce the LEP constraints on
the lightest CP-even Higgs $H_{1}$ and LHC constraints on the heavier
scalars.  The near mass degeneracy of $H_{3},\,A,\,H^{\pm}$ and
$\Delta^{\pm\pm}$ ensures that the oblique parameters are not
affected.  In analogy to the type-II seesaw model with explicit \lnv
we expect that, because of $v_{3}<10^{-4}~\g$, the doubly-charged
scalar predominantly decays into same sign
dileptons~\cite{Sugiyama:2012yw,delAguila:2013mia,Han:2015hba} and
that $m_{\Delta^{\pm\pm}}=500~\g$ is consistent with current
experimental data, see subsection~\ref{chargedhiggs}.\\

\begin{figure}[H]
\vspace{-5pt}
\centering
  \includegraphics[width=0.6\textwidth]{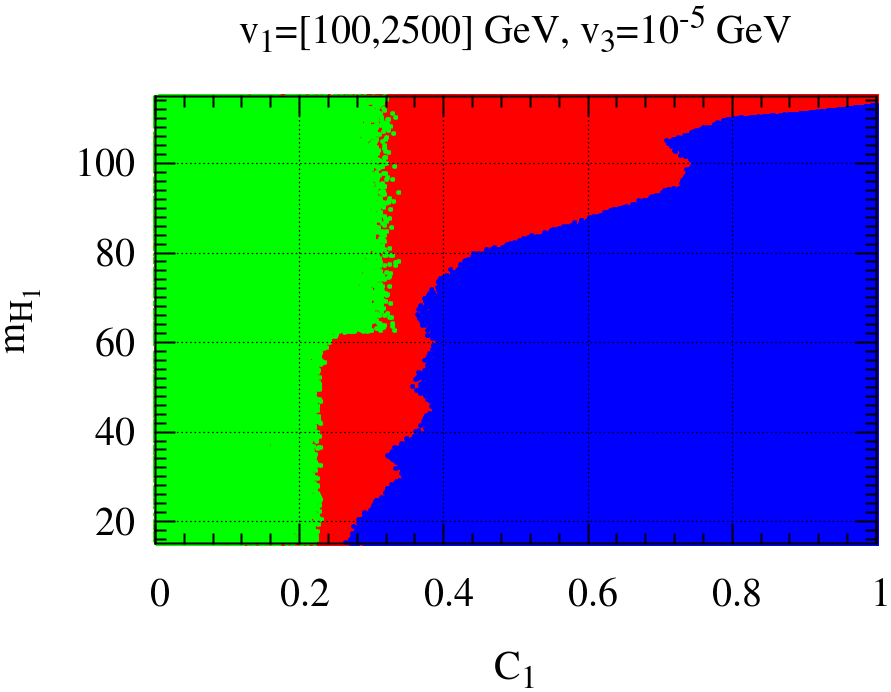}\quad
\vspace{-5pt}
  \caption{Analysis (i). The mass of the lightest 
  CP-even scalar as a function of the absolute value of its 
coupling to \sm states. The blue region corresponds to the LEP 
exclusion region and the green(red) one is the LHC 
allowed(exclusion) region.
\label{fig:A1}}
\end{figure}

We show in Fig.~\ref{fig:A1} the mass of the lightest CP-even scalar
as a function of the absolute value of its coupling to the \sm states,
$|C_{1}|$ in eq.~(\ref{htosm}). The blue region corresponds to the LEP
exclusion region and the green(red) one is the LHC allowed(exclusion)
region provided by the signal strengths $0.8<\mu_{XX}<1.2$.\\

\begin{figure}[H]
\vspace{-0pt}
  \centering
  \includegraphics[width=0.45\textwidth]{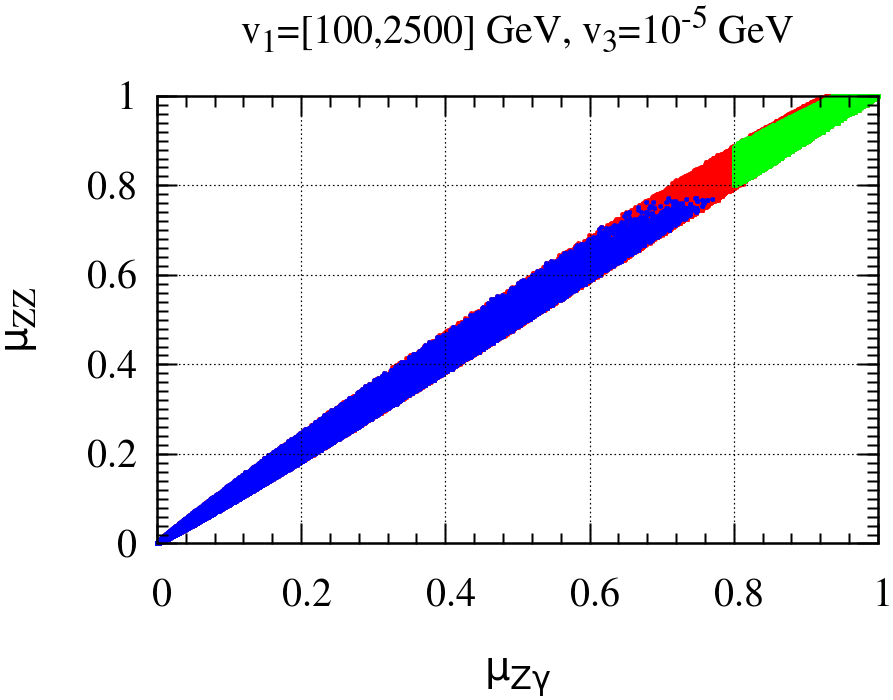} \quad
  \includegraphics[width=0.45\textwidth]{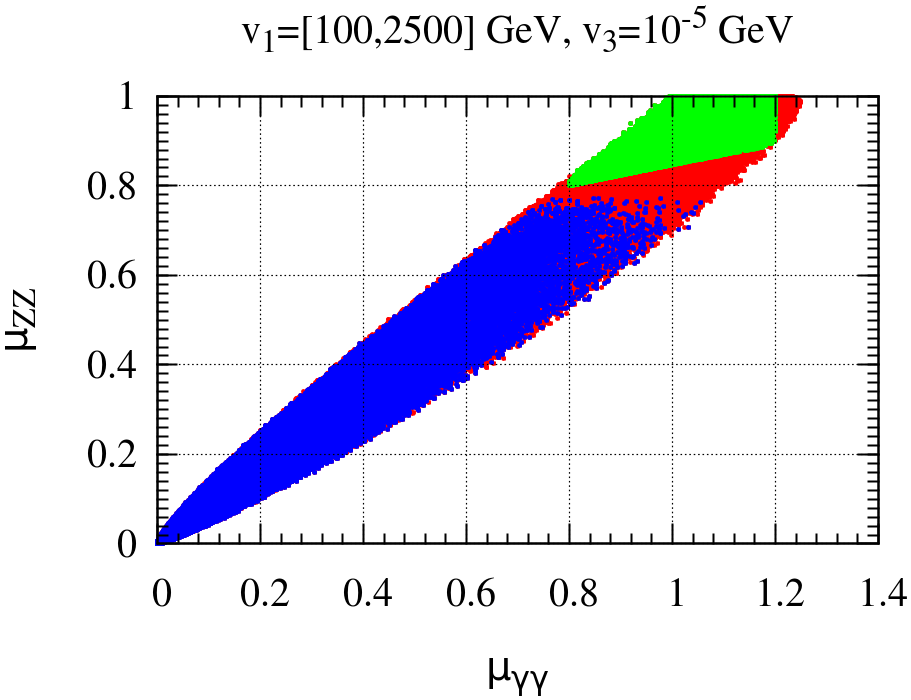} \\
\vspace{-5pt}
  \caption{Analysis (i). On the left, we show 
  correlation between $\mu_{ZZ}$ and $\mu_{Z\gamma}$. On the right, 
  correlation between $\mu_{ZZ}$ and $\mu_{\gamma\gamma}$.  
  The color code as in Fig.~\ref{fig:A1}.
  \label{fig:A2}}
\end{figure}

The presence of light charged scalars can enhance significantly the
diphoton channel of the Higgs~\cite{Akeroyd:2012ms}. Fig.~\ref{fig:A2}
shows the correlation between $\mu_{ZZ}$ and
$\mu_{\gaga}$($\mu_{Z\gamma}$) on the left(right) with
$\mu_{\gaga}\lesssim1.2$ for charged Higgs bosons of $500\g$.
The correlation between the signal strength $\mu_{ZZ}$ and the signal
strengths $\mu_{\gamma\gamma}$ and $\mu_{Z\gamma}$ is shown in
Fig.~\ref{fig:A2}. Note that the former may exceed one due to the new
contributions of the singly and doubly charged Higgs bosons.

\begin{figure}[H]
\vspace{-5pt}
  \centering
  \includegraphics[width=0.45\textwidth]{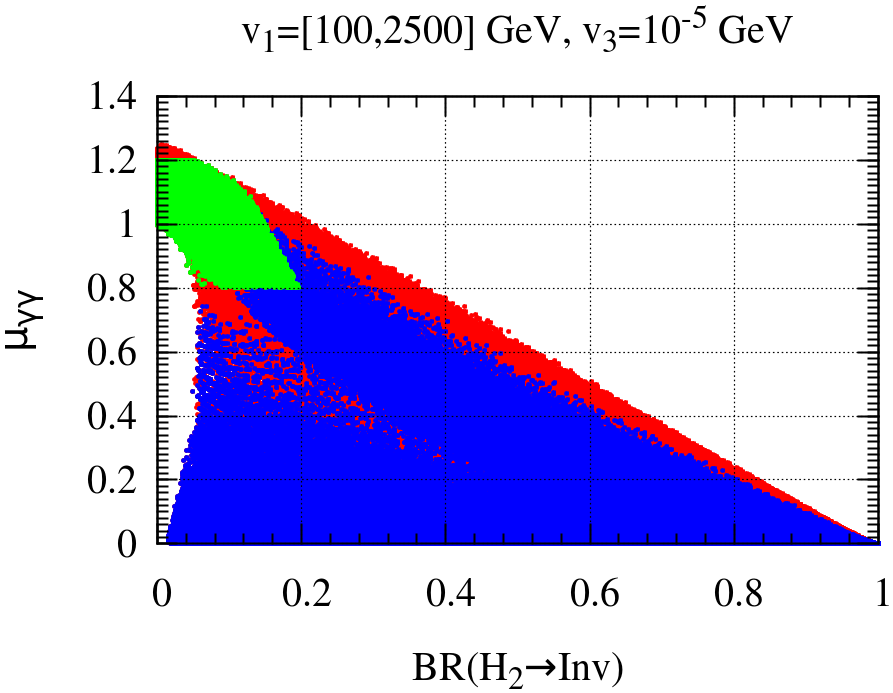} \quad 
 \includegraphics[width=0.45\textwidth]{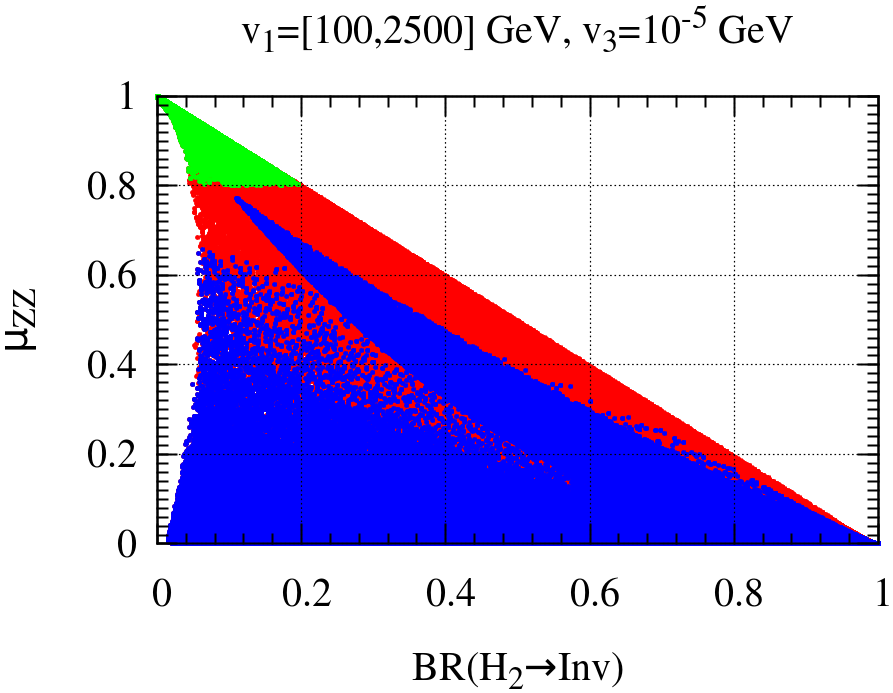}\\
\vspace{-5pt}
  \caption{Analysis (i). On the left: the signal 
  strength $\mu_{\gaga}$ versus $\text{BR}(H_{2}\to\text{Inv})$. 
  On the right: $\mu_{ZZ}$ versus $\text{BR}(H_{2}\to\text{Inv})$. 
  The color code as in Fig.~\ref{fig:A1}. \label{fig:A3}}
\end{figure}

The invisible decays of the Higgs bosons, characteristic of the model,
turn out to be correlated to the visible channels, represented in
terms of the signal strengths, as shown in Fig.~\ref{fig:A3}.  Note
that the upper bound on the invisible decays of a Higgs boson with a
mass of 125 GeV has been found to be $\text{BR}
(H_{2}\to\text{Inv})\lesssim0.2$. This limit is stronger than those
provided by the ATLAS~\cite{Aad:2015txa} and the
CMS~\cite{Chatrchyan:2014tja} collaborations\footnote{ The ATLAS
  collaboration has set an upper bound on the BR($H\to\text{Inv}$) at
  0.28 while the CMS collaboration reported that the observed
  (expected) upper limit on the invisible branching ratio is
  0.58(0.44), both results at 95\% C.L.}.\\

\begin{figure}[H]
\vspace{-5pt}
  \centering
  \includegraphics[width=0.45\textwidth]{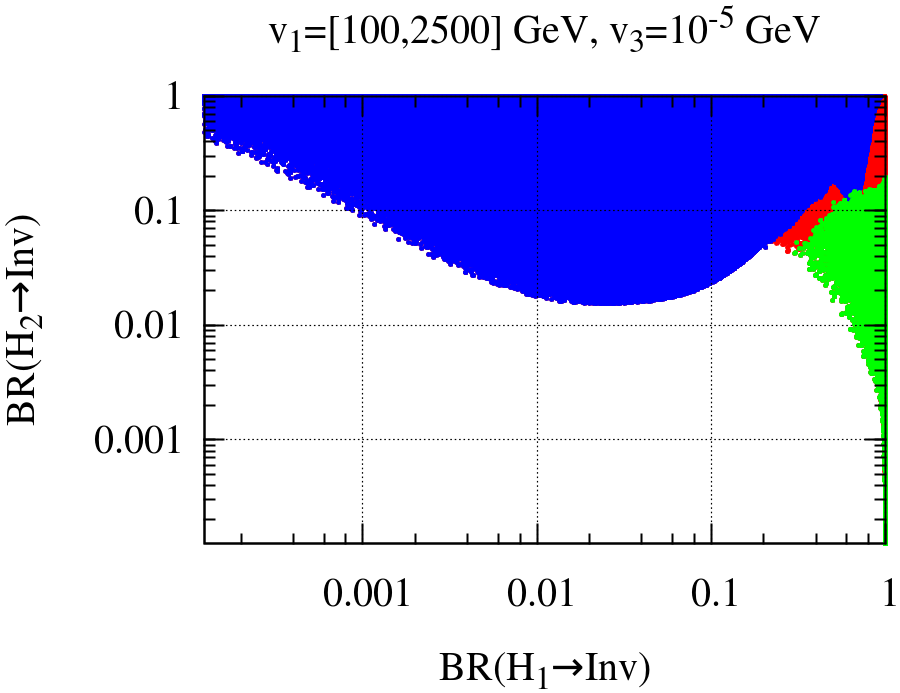} \quad
\vspace{-5pt}
  \caption{Analysis (i). Correlation between the invible branchings 
  $\text{BR}(H_{2}\to\text{Inv})$ and $\text{BR}(H_{1}\to\text{Inv})$. 
  The color code as in Fig.~\ref{fig:A1}.
  \label{fig:A4}}
\end{figure}

In Fig.~\ref{fig:A4} we depict the correlation between the invisible
branching ratios of $H_2$ with the one of the lightest scalar boson
$H_1$. And, as can be seen, $H_{1}$ can decay 100\% into the invisible
channel (majorons).


Finally, as we have mentioned we obtained that the reduced coupling of
$H_{3}$ to the \sm states is $C_{3}\sim \mathcal{O}(10^{-7})$ so that
it is basically decoupled. As a result its invisible branching
is essentially unconstrained, $10^{-5}\lesssim\text{BR}(H_{3}
\to\text{Inv})\leq1$. On the other hand we find that the constraint
coming from the LHC on the pseudo-scalar $A$ with a mass of $500\g$
is automatically satisfied as well, since from the LHC, $\sigma(gg\to
A)\text{BR}(A\to ZH_{2})\text{BR}(H_{2}\to \tau\tau) \lesssim10^{-2}$
while for $m_A=500\g$ we obtain $\sigma(gg\to A)
\text{BR}(A\to ZH_{2})\text{BR}(H_{2}\to \tau\tau)\lesssim10^{-15}$.\\

\subsection{Analysis (ii)}
\label{sec:analysis-ii}

\begin{figure}[H]
\vspace{-5pt}
  \centering
  \includegraphics[width=0.45\textwidth]{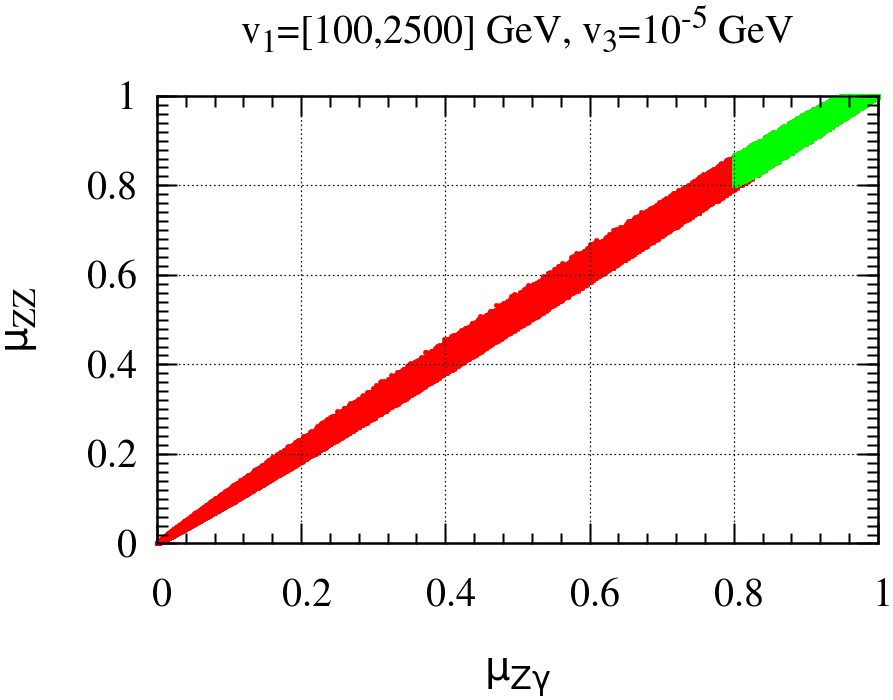} \quad
  \includegraphics[width=0.45\textwidth]{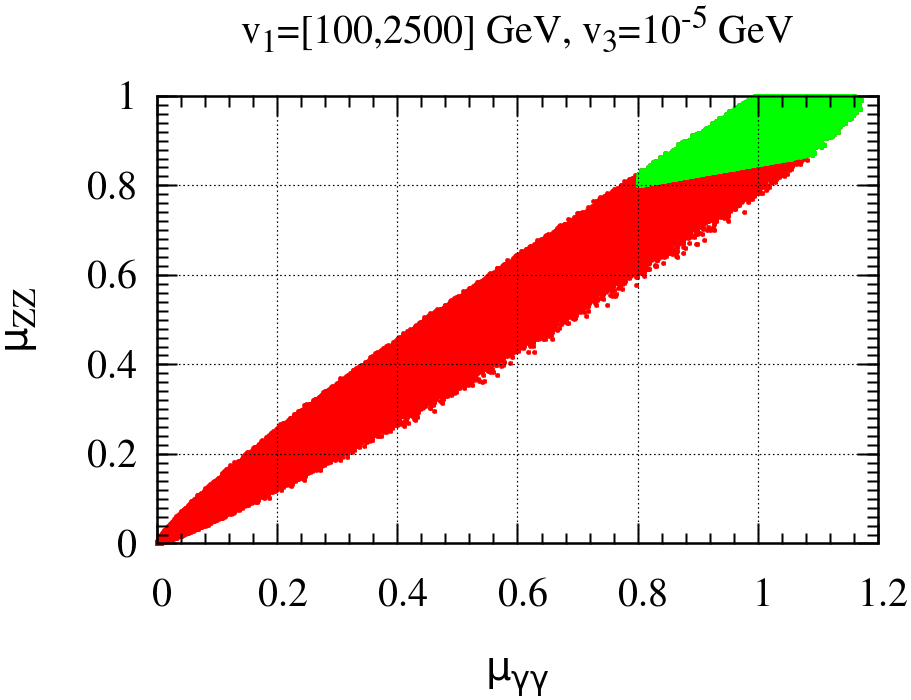} \\
\vspace{-5pt}
  \caption{Analysis (ii). On the left, $\mu_{ZZ}$ 
  versus $\mu_{Z\gamma}$. On the right, $\mu_{ZZ}$ versus 
  $\mu_{\gamma\gamma}$. The allowed region (in green) is the 
  region inside the range $\mu_{XX} = 1.0^{+0.2}_{-0.2}$ while 
  the forbidden one (in red) is the one outside that range.
  \label{fig:B1}}
\end{figure}

We now turn to the other case of interest, namely
\bea
&m_{H_{1}}=125\g,\ \ m_{H_{2}}=[150,500]~\text{GeV},\ \ 
m_{H_{3}}\simeq m_{A}\simeq m_{H^{\pm}}\simeq m_{\Delta^{\pm\pm}}=600\g,\nn
\eea
with $v_{3}=10^{-5}~\g$, as before. Now  we scanned over
\bea
&v_{1}\in[100,2500]~\g,\ \ \alpha_{12}\in[0,\pi]\ \ \text{and}\ \ \alpha_{13,23}=
\delta_\alpha\,(\pi-\delta_\alpha) 
\eea
where $0\leq\delta_\alpha<0.1$. As we already mentioned in this case
we only have to take into account the constraints coming from 
Run 1 of the LHC at $8~\text{TeV}$, see Table~\ref{tab:1}.
In practice we assume $\mu_{XX}=1.0^{+0.2}_{-0.2}$. We show in 
Fig.~\ref{fig:B1} the correlation between $\mu_{ZZ}$ and 
$\mu_{\gaga}$($\mu_{Z\gamma}$) on the left(right). As before, 
the allowed region is in green while the forbidden one is in red. We
can see that $\mu_{\gaga}\lesssim1.2$ for $m_{H^{\pm}}\simeq 
m_{\Delta^{\pm\pm}}=600\g$.\\

\begin{figure}[H]
\vspace{-5pt}
  \centering
  \includegraphics[width=0.45\textwidth]{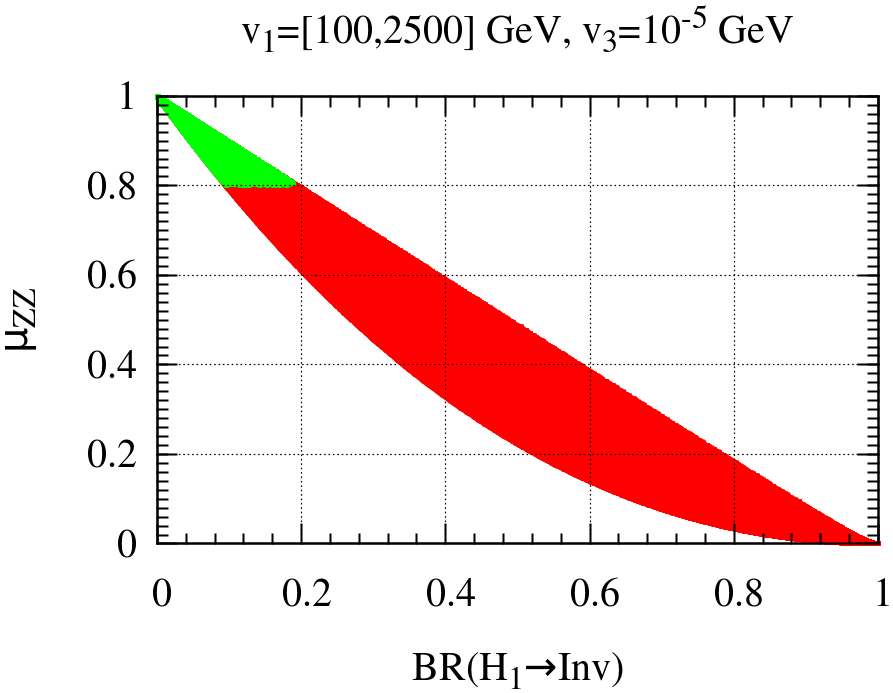}\quad 
  \includegraphics[width=0.45\textwidth]{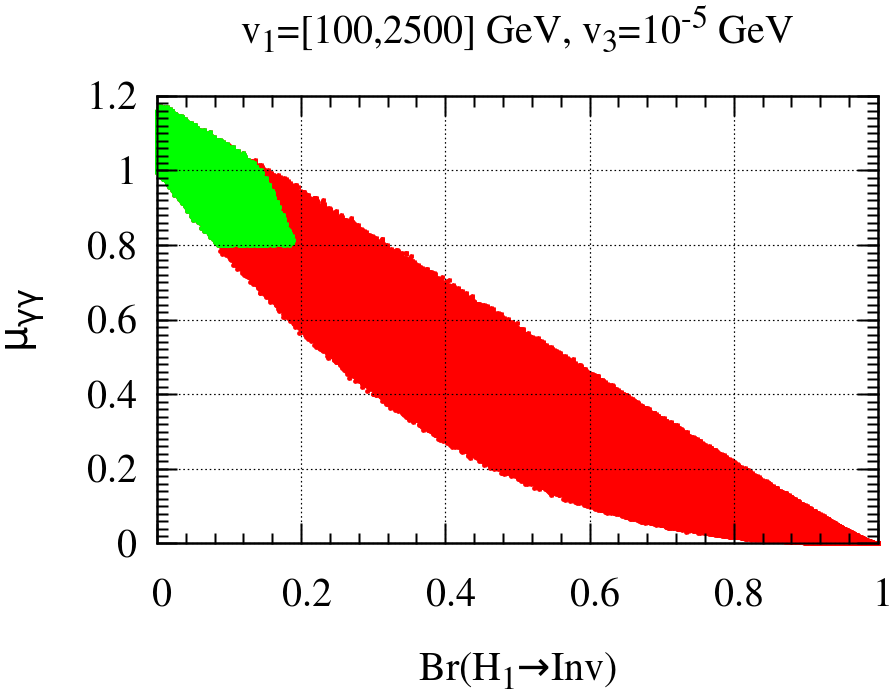} \\
\vspace{-5pt}
  \caption{Analysis (ii). On the left: the signal 
  strength $\mu_{ZZ}$ versus
  $\text{Br}(H_{1}\to\text{Inv})$. On the right: $\mu_{\gaga}$ versus
  $\text{Br}(H_{1}\to\text{Inv})$. The color
  code as in Fig.~\ref{fig:B1} \label{fig:B2}}
\end{figure}

\begin{figure}[H]
  \centering
  \includegraphics[width=0.45\textwidth]{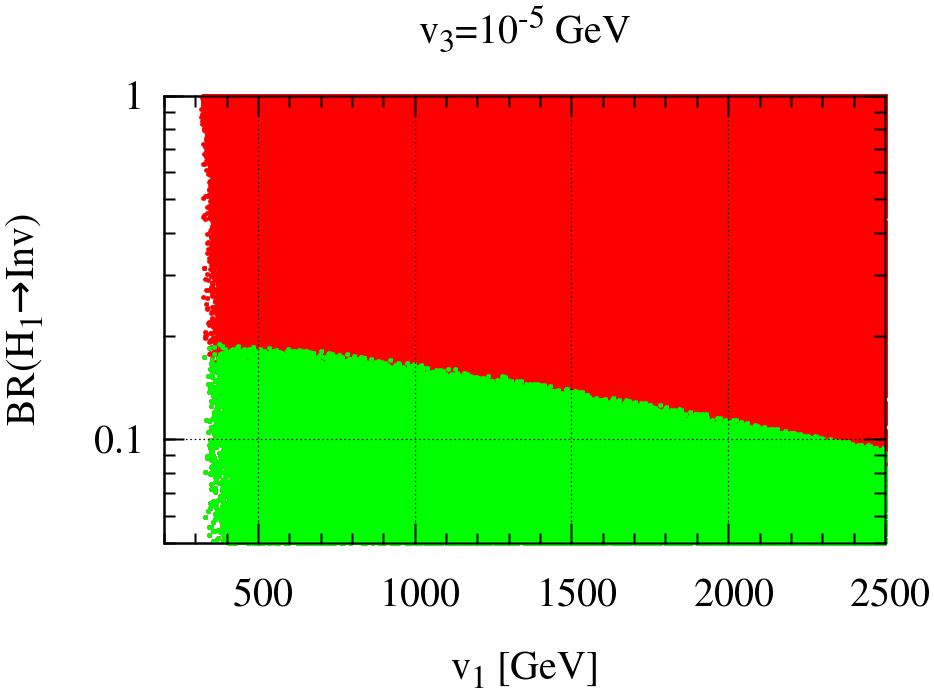}\quad 
  \includegraphics[width=0.45\textwidth]{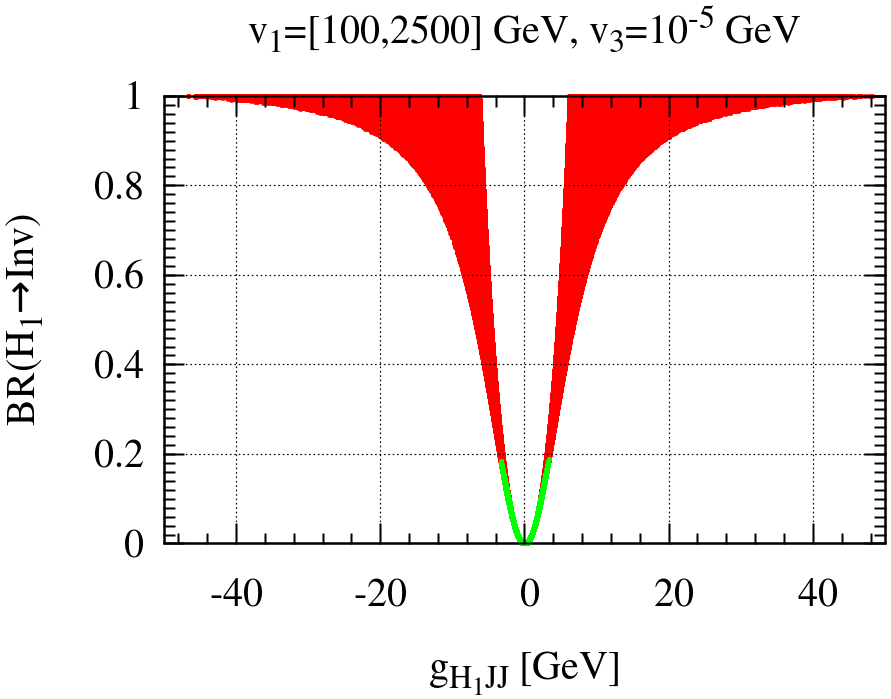} \\
  \vspace{-5pt}
  \caption{Analysis (ii). $\text{BR}(H_{1}\to\text{Inv})$ versus
  $v_1$ (on the left) and $\text{BR}(H_{1}\to\text{Inv})$ versus
  the Higgs-majoron coupling $g_{H_{1}JJ}$ (on the right). The color
  code as in Fig.~\ref{fig:B1}. 
  \label{fig:B3}}
\end{figure}

\begin{figure}[H]
\vspace{-5pt}
  \centering
  \includegraphics[width=0.45\textwidth]{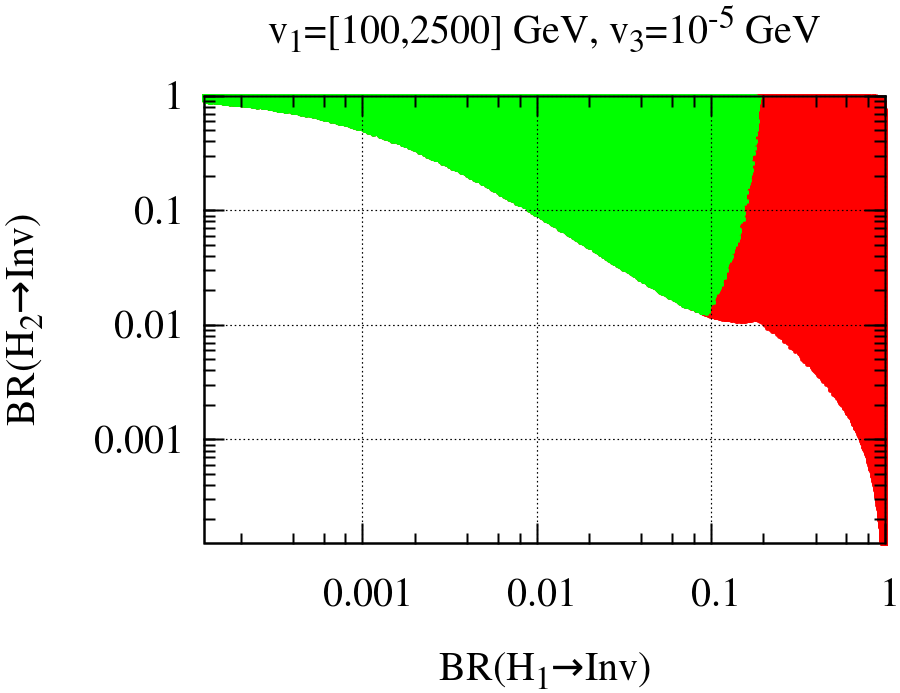} \quad
\vspace{-5pt}
  \caption{Analysis (ii). Correlation between $\text{BR}(H_{2}\to\text{Inv})$ 
  and $\text{BR}(H_{1}\to\text{Inv})$. The color
  code as in Fig.~\ref{fig:B1}.
  \label{fig:B4}}
\end{figure}

On the left(right) of Fig.~\ref{fig:B2} is depicted the 
correlation between the signal strength $\mu_{ZZ}$ ($\mu_{\gaga}$)
and the branching ratio of the channel $H_{1}\to JJ$. We
can see in Figs.~\ref{fig:B2}-\ref{fig:B4} that 
$\text{BR}(H_{1}\to\text{Inv})\lesssim 0.2$.
One can see from Fig.~\ref{fig:B3} that $\text{BR}(H_{1}\to\text{Inv})
\lesssim 0.1$  for $v_{1}\gtrsim2500\g$.
In this case we find that eq.~\ref{ghJJ} (for $\alpha_{13,23}\sim0(\pi)$
and $v_3\ll v_{1},v_2$) at leading order is given by,
\begin{equation}
 g_{H_{1}JJ}\sim \frac{\cos\alpha_{12}}{v_1}m_{H_1}^2,
\end{equation}
where $m_{H_{1}}=125\g$. $\text{BR}(H_{1}\to\text{Inv})$ versus
the Higgs-majoron coupling $g_{H_{1}JJ}$ is shown on the right of
Fig.~\ref{fig:B3}. 
Note also from the left panel in Fig.~\ref{fig:B3} that
$\text{BR}(H_{1}\to\text{Inv})$ is anti-correlated with $v_1$, as
expected.

In Fig.~\ref{fig:B4} we show the correlation between the invisible
branching ratio of $H_{2}$ (the Higgs with a mass in the range
$150\g <m_{H_{2}}<500\g$) and the one of $H_{1}$.

\begin{figure}[H]
\centering
  \includegraphics[width=0.45\textwidth]{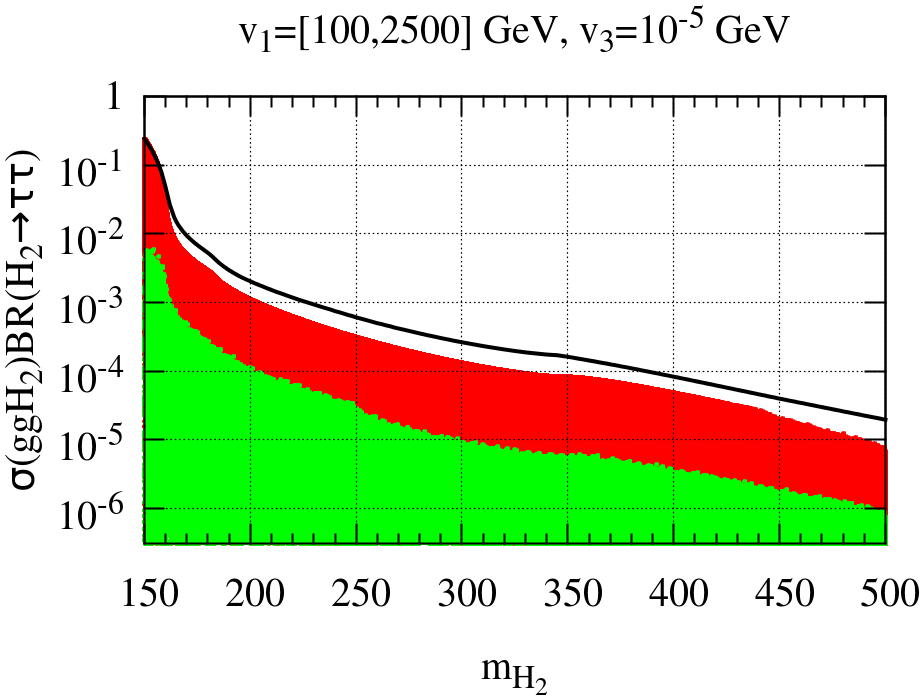}\quad 
  \includegraphics[width=0.45\textwidth]{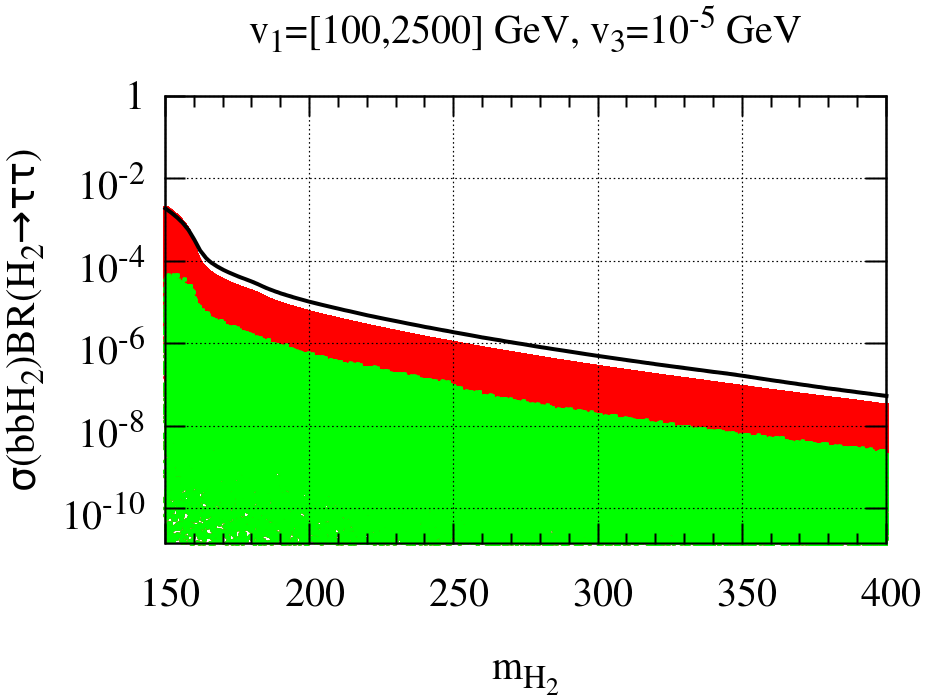} \\
  \vspace{-5pt}
   \caption{Analysis (ii). On the top right (left) 
   $\sigma(ggH_{2})\text{BR}(H_{2}\to \tau\tau)$ ($\sigma(bbH_{2})
   \text{BR}(H_{2}\to\tau\tau)$) versus the mass of $H_{2}$. 
  \label{fig:B5}}
 \end{figure}

We have verified that the LHC constraints on the heavy scalars
($H_{2}$, $H_{3}$ and $A$) are all satisfied. As an example, the
reader can convince her/himself by looking at Fig.~\ref{fig:B5} that
$H_{2}$ easily passes the restriction stemming from
$\sigma(ggH_{2})\text{BR} (H_{2}\to\tau\tau)$ (top left) and/or
$\sigma(bbH_{2})\text{BR}(H_{2}\to\tau\tau)$ (top right).
The black continuous lines on those plots represent the experimental
results from Run 1 of the CMS experiment \cite{Khachatryan:2014wca}.
We also found that the square of the reduced coupling of $H_2$ to the
Standard Model states is $C_{2}^2\lesssim0.1$ for
$m_{H_{2}}=[150,500]~\g$.  
Then, one finds that the experimental upper bounds set by the search
for a heavy Higgs in the $H\to WW$ and $H\to ZZ$ decay channels in
\cite{Khachatryan:2014jba,Khachatryan:2015cwa} are automatically
fulfilled.  However, improved sensitivities expected from Run 2 may
provide a meaningful probe of the theoretically consistent region,
depicted in green.

 Also in this case, $H_{3}$ is decoupled, so the restrictions on
 $H_{3}$ and the massive pseudoscalar $A$ are automatically fulfilled.

\section{Conclusions}

In this paper we have presented the main features of the electroweak
symmetry breaking sector of the simplest type-II seesaw model with
spontaneous violation of lepton number. 
The Higgs sector has two characteristic features: a) the existence of a
(nearly) massless Nambu-Goldstone boson and b) all neutral CP-even and
CP-odd, as well as singly and doubly-charged scalar bosons coming
mainly from the triplet are very close in mass, as illustrated in
Fig.~\ref{fig:ms} of appendix~\ref{massspec}. However, one extra  CP-even
state, namely $H_2$ coming from a doublet-singlet mixture can be light.
After reviewing the ``theoretical'' and experimental restrictions
which apply on the Higgs sector, we have studied the sensitivities of
the searches for Higgs bosons at the ongoing ATLAS/CMS experiments,
including not only the new contributions to \sm decay channels, but
also the novel Higgs decays to majorons.
For these we have considered two cases, when the 125 GeV state found
at CERN is either (i) the second-to-lightest or (ii) the lightest
CP-even scalar boson.
For case (i), we have enforced the constraints coming from LEP-II data
on the lightest CP-even scalar coupling to the \sm states and those
coming from the LHC Run-1 on the heavier scalars.  
In case (ii), only the constraints coming from the LHC must be taken
into account.
Such ``invisible'' Higgs boson decays give rise to missing momentum
events. We have found that the experimental results from Run 1 on the
search for a heavy Higgs in the $H\to WW$ and $H\to ZZ$ decay channels
are automatically fulfilled. However, improved sensitivities expected
from Run 2 may provide a meaningful probe of this scenario.
In short we have discussed how the neutrino mass generation scenario
not only suggests the need to reconsider the physics of electroweak
symmetry breaking from a new perspective, but also provides a new
theoretically consistent and experimentally viable paradigm.

\section{Acknowledgments}

Work supported by the Spanish grants FPA2014-58183-P, Multidark
CSD2009-00064 and SEV-2014-0398 (MINECO), and PROMETEOII/2014/084
(Generalitat Valenciana).  C. B. thanks Departamento de Física and
CFTP, Instituto Superior Técnico, Universidade de Lisboa, for its
hospitality while part of this work was carried out.  J.C.R. is also
support in part by the Portuguese Funda\c{c}\~ao para a Ci\^encia e
Tecnologia (FCT) under contracts UID/FIS/00777/2013 and
CERN/FIS-NUC/0010/2015, which are partially funded through POCTI
(FEDER), COMPETE, QREN and the EU.

\newpage
\appendix

\section{Higgs boson mass spectrum}
\label{massspec}

 \begin{figure}[H]
\vspace{-5pt}
  \centering
  \includegraphics[width=0.7\textwidth]{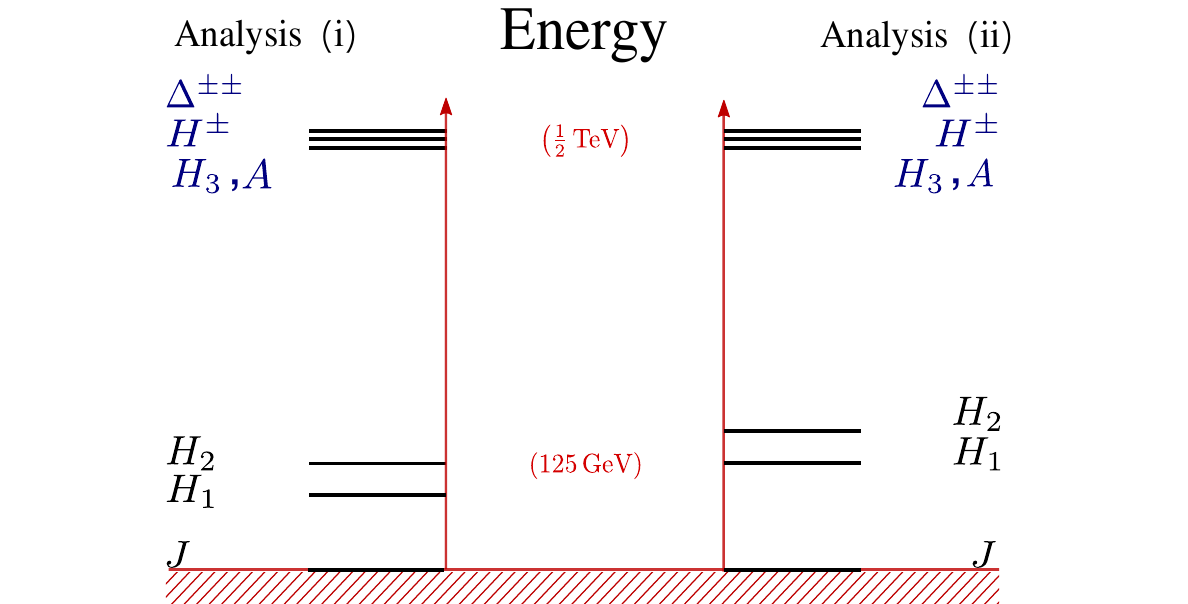} \quad
\vspace{-5pt}
  \caption{Type-II seesaw Higgs boson mass spectrum
  \label{fig:ms}}
\end{figure}

\begin{table}[H]
\begin{centering}
\begin{tabular}{||c|c|c||}
\hline 
Mass eigenstate $\phi$ & Mass squared $m_{\phi}^{2}$ & Composition\tabularnewline
\hline
$H_{i}$\,$(i=1,2,3)$& $m_{i}^2$   & $\mathcal{O}^{R}_{i1}R_{1}+\mathcal{O}^{R}_{i2}R_{2}+\mathcal{O}^{R}_{i3}R_{3}$\tabularnewline
$J$                  & $0$         & $\mathcal{O}^{I}_{11}I_{1}+\mathcal{O}^{I}_{12}I_{2}+\mathcal{O}^{I}_{13}I_{3}$\tabularnewline
$G^{0}$              & $0$         & $\mathcal{O}^{I}_{22}I_{2}+\mathcal{O}^{I}_{23}I_{3}$\tabularnewline
$A$                  & $\kappa\left(\frac{v_2^2 v_1^2 +v_2^2v_3^2 + 4v_3^2 v_1^2}{2v_3 v_1}\right)$& $\mathcal{O}^{I}_{31}I_{1}+\mathcal{O}^{I}_{32}I_{2}+\mathcal{O}^{I}_{33}I_{3}$\tabularnewline
$G^{\pm}$            & $0$         & $c_{\pm}\phi^{\pm}+s_{\pm}\Delta^{\pm}$\tabularnewline
$H^{\pm}$            & $\frac{1}{4v_3}(2\kappa v_1-\lambda_5 v_3)(v_2^2+2v_3^2)$ & $-s_{\pm}\phi^{\pm}+c_{\pm}\Delta^{\pm}$\tabularnewline
$\Delta^{\pm\pm}$    & $\frac{1}{2v_3}(\kappa v_1
v_2^2- 2\lambda_4 v_3^3-\lambda_5 v_2^2 v_3)$         & $\Delta^{\pm\pm}$\tabularnewline
\bottomrule
\end{tabular}
\par\end{centering}
\protect\caption{\label{tab:ms}Scalar mass eigenstates in the model.
$c_{\pm}=v_{2}/\sqrt{v_{2}^{2}+2v_{3}^{2}}$, $s_{\pm}=\sqrt{2}v_{3}/\sqrt{v_{2}^{2}+2v_{3}^{2}}$}
\end{table}

\newpage
\section{Loop functions}
\label{loopfuncs}

The one-loop function $f(\tau)$ used in eq.~(\ref{ffactor1})
is given by,
\begin{eqnarray}
f(\tau) =
\left\{
	\begin{array}{ll}
\arcsin^{2}(1/\sqrt{\tau})  & \text{if } \tau\geq1 \\
-\frac{1}{4}\left[\log\left(\frac{1+\sqrt{1-\tau}}{1-\sqrt{1-\tau}}\right)-i\pi\right]^2  & \mbox{if } \tau<1
	\end{array}
\right.
\end{eqnarray}\\

The functions $C^{b}_0$ and $\Delta B^b_{0}$ are given in 
terms of the Passarino-Veltman functions \cite{Passarino:1978jh},
\begin{eqnarray}
C^{b}_0&=&C_0(m_Z^2,0,m_{H_a}^2,m_b^2,m_b^2,m_b^2)=-\frac{1}{m_b^2}I_2(\tau_b,\lambda_b),
\nn\\
\Delta B^b_{0}&=& B_0(m_{H_a}^2,m_b^2,m_b^2)-B_0(m_{Z}^2,m_b^2,m_b^2)
\notag\\
&=&-\frac{m_{H_a}^2-m_Z^2}{m_Z^2}-\frac{(m_{H_a}^2-m_Z^2)^2}{2m_a^2 m_Z^2}I_1(\tau_b,\lambda_b)
+ 2\frac{m_{H_a}^2-m_Z^2}{m_Z^2}I_2(\tau_b,\lambda_b)
\end{eqnarray}
where $\lambda_b=4m_b^2/m_{H_a}^2$,
\begin{eqnarray}
I_{1}(\tau,\lambda)&=&\frac{\tau\lambda}{2(\tau-\lambda)}+\frac{\tau^{2}\lambda^{2}}{2(\tau-\lambda)^{2}}
\left[f(\tau)-f(\lambda)\right]+\frac{\tau^{2}\lambda}{(\tau-\lambda)^{2}}\left[g(\tau)-g(\lambda)\right],
\notag\\
I_{2}(\tau,\lambda)&=&-\frac{\tau\lambda}{2(\tau-\lambda)}\left[f(\tau)-f(\lambda)\right],
\end{eqnarray}
and
\begin{eqnarray}
g(\tau) =
\left\{
	\begin{array}{ll}
\sqrt{\tau-1}\arcsin\sqrt{\tau}  & \mbox{for } \tau\geq1 \\
\frac{1}{2}\sqrt{1-\tau}\left(\log\frac{1+\sqrt{1-\tau}}{1-\sqrt{1-\tau}}-i\pi\right)  & \mbox{if } \tau<1
	\end{array}
\right.
\end{eqnarray}

\section{Higgs boson couplings}\label{gcps}

\begin{table}[H]
\begin{tiny}
 \begin{tabular}{|c|c|c|}
  \hline
               &             Vertex                 & Gauge Coupling \\
  \hline
  1            & $H_{1}W^{+}_{\mu}W_{\nu}^{-}$      & $i\frac{g^{2}}{2}(\mathcal{O}^{R}_{12} v_{2}+2 \mathcal{O}^{R}_{13} v_{3})\gmn$\\
  2         & $H_{2}W^{+}_{\mu}W_{\nu}^{-}$      & $i\frac{g^{2}}{2}(\mathcal{O}^{R}_{22} v_{2}+2 \mathcal{O}^{R}_{23} v_{3})\gmn$\\
  3            & $H_{3}W^{+}_{\mu}W_{\nu}^{-}$      & $i\frac{g^{2}}{2}(\mathcal{O}^{R}_{32} v_{2}+2 \mathcal{O}^{R}_{33} v_{3})\gmn$\\
  4            & $\Delta^{\pm\pm}W^{\mp}_{\mu}W_{\nu}^{\mp}$& $i2g^{2}\frac{v_{3}}{\sqrt{2}}g_{\mu\nu}$\\
  5            & $H^{\pm} W^{\mp}_{\mu}Z_{\nu}$         & $i\frac{g^{2}}{c_{W}}\frac{c_{\pm}v_{3}}{\sqrt{2}}g_{\mu\nu}$ \\
  6         & $G^{\pm}W^{\mp}_{\mu}Z_{\nu}$          & $i\frac{g^{2}}{c_{W}}\left(\frac{v_{2}}{2}s_{W}^{2}c_{\pm}+\frac{v_{3}}{\sqrt{2}}(1+s_{W}^{2})s_{\pm}\right) g_{\mu\nu}$\\                         
  7         & $G^{\pm}W^{\mp}_{\mu}A_{\nu}$          & $-i e m_{W}g_{\mu\nu}$\\
  8            &$\Delta^{++}\Delta^{--}W_{\mu}^{+}W_{\nu}^{-}$& $ig^{2}\gmn$ \\
  9            & $H^{+}H^{-}W_{\mu}^{+}W_{\nu}^{-}$ & $i\frac{g^{2}}{2}(1+3c_{\pm}^{2})\gmn$\\
  10        & $G^{+}G^{-}W_{\mu}^{+}W_{\nu}^{-}$ & $i\frac{g^{2}}{2}(1+3s_{\pm}^{2})\gmn$ \\
  11           & $H_{1}H_{1}W_{\mu}^{+}W_{\nu}^{-}$ & $i\frac{g^{2}}{2}(\mathcal{O}^{R\,2}_{12}+2\mathcal{O}^{R\,2}_{13})\gmn$ \\
  12        & $H_{2}H_{2}W_{\mu}^{+}W_{\nu}^{-}$ & $i\frac{g^{2}}{2}(\mathcal{O}^{R\,2}_{22}+2\mathcal{O}^{R\,2}_{23})\gmn$ \\
  13           & $H_{3}H_{3}W_{\mu}^{+}W_{\nu}^{-}$ & $i\frac{g^{2}}{2}(\mathcal{O}^{R\,2}_{32}+2\mathcal{O}^{R\,2}_{33})\gmn$ \\
  14          & $JJ W_{\mu}^{+}W_{\nu}^{-}$        & $i\frac{g^{2}}{2}(\mathcal{O}^{I\,2}_{12}+2\mathcal{O}^{I\,2}_{13})\gmn$  \\
  15        & $G^{0}G^{0}W_{\mu}^{+}W_{\nu}^{-}$ & $i\frac{g^{2}}{2}(\mathcal{O}^{I\,2}_{22}+2\mathcal{O}^{I\,2}_{23})\gmn$  \\
  16           & $A A W_{\mu}^{+}W_{\nu}^{-}$       & $i\frac{g^{2}}{2}(\mathcal{O}^{I\,2}_{32}+2\mathcal{O}^{I\,2}_{33})\gmn$  \\
  17           & $\Delta^{\pm\pm}H^{\mp} W_{\nu}^{\mp}$     & $\mp igc_{\pm}\ppm$ \\ 
  18           & $\Delta^{\pm\pm}G^{\mp} W_{\nu}^{\mp}$     & $\mp igs_{\pm}\ppm$ \\
  19           & $H_{1}H^{\pm}W^{\mp}_{\mu}$           & $\pm i\frac{g}{2}(s_{\pm}\mathcal{O}^{R}_{12}-\sqrt{2}c_{\pm}\mathcal{O}^{R}_{13})\ppm$\\
  20           & $H_{2}H^{\pm}W^{\mp}_{\mu}$           & $\pm i\frac{g}{2}(s_{\pm}\mathcal{O}^{R}_{22}-\sqrt{2}c_{\pm}\mathcal{O}^{R}_{23})\ppm$ \\
  21           & $H_{3}H^{\pm}W^{\mp}_{\mu}$           & $\pm i\frac{g}{2}(s_{\pm}\mathcal{O}^{R}_{32}-\sqrt{2}c_{\pm}\mathcal{O}^{R}_{33})\ppm$ \\
  22           & $H^{\pm}J W^{\mp}_{\mu}$               & $\frac{g}{2}(s_{\pm}\mathcal{O}^{I}_{12}+\sqrt{2}c_{\pm}\mathcal{O}^{I}_{13})\ppm$ \\
  23           & $G^{0}H^{\pm}W^{\mp}_{\mu}$           & $-\frac{g}{2}(s_{\pm}\mathcal{O}^{I}_{22}+\sqrt{2}c_{\pm}\mathcal{O}^{I}_{23})\ppm$ \\
  24           & $AH^\pm W^{\mp}_{\mu}$               & $-\frac{g}{2}(s_{\pm}\mathcal{O}^{I}_{32}+\sqrt{2}c_{\pm}\mathcal{O}^{I}_{33})\ppm$ \\
  25           & $G^{\pm}H_{1} W^{\mp}_{\mu}$           & $\pm i\frac{g}{2}(c_{\pm}\mathcal{O}^{R}_{12}+\sqrt{2}s_{\pm}\mathcal{O}^{R}_{13})\ppm$\\
  26        & $G^{\pm}H_{2} W^{\mp}_{\mu}$           & $\pm i\frac{g}{2}(c_{\pm}\mathcal{O}^{R}_{22}+\sqrt{2}s_{\pm}\mathcal{O}^{R}_{23})\ppm$ \\ 
  27           & $G^{\pm}H_{3} W^{\mp}_{\mu}$           & $\pm i\frac{g}{2}(c_{\pm}\mathcal{O}^{R}_{32}+\sqrt{2}s_{\pm}\mathcal{O}^{R}_{33})\ppm$ \\
  28           & $G^{\pm}J W^{\mp}_{\mu}$               & $-\frac{g}{2}(c_{\pm}\mathcal{O}^{I}_{12}-\sqrt{2}s_{\pm}\mathcal{O}^{I}_{13})\ppm$ \\
  29        & $G^{0}G^\pm W^{-}_{\mu}$            & $\frac{g}{2}(c_{\pm}\mathcal{O}^{I}_{22}-\sqrt{2}s_{\pm}\mathcal{O}^{I}_{23})\ppm$\\
  30           & $A G^\pm W^{-}_{\mu}$               & $\frac{g}{2}(c_{\pm}\mathcal{O}^{I}_{32}-\sqrt{2}s_{\pm}\mathcal{O}^{I}_{33})\ppm$ \\ 
  \hline
\end{tabular}
\ \
\begin{tabular}{|c|c|c|}
  \hline
               &             Vertex                 & Gauge Coupling \\
  \hline
  31           & $H_{1}Z_{\mu}Z_{\nu}$              & $i\frac{g^{2}}{2c_{W}^{2}}(\mathcal{O}^{R}_{12}v_{2}+4 \mathcal{O}^{R}_{13}v_{3})\gmn$\\
  32        & $H_{2}Z_{\mu}Z_{\nu}$              & $i\frac{g^{2}}{2c_{W}^{2}}(\mathcal{O}^{R}_{22}v_{2}+4 \mathcal{O}^{R}_{23}v_{3})\gmn$\\ 
  33           & $H_{3}Z_{\mu}Z_{\nu}$              & $i\frac{g^{2}}{2c_{W}^{2}}(\mathcal{O}^{R}_{32}v_{2}+4 \mathcal{O}^{R}_{33}v_{3})\gmn$\\
  34           & $\Delta^{++}\Delta^{--}Z_{\mu}Z_{\nu}$& $i\frac{2g^{2}}{c_{W}^{2}}(c_{W}^{2}-s_{W}^{2})^2 \gmn$ \\
  35           & $H^{+}H^{-}Z_{\mu}Z_{\nu}$         & $i\frac{g^{2}}{2c_{W}^{2}}(s_{\pm}^{2}(c_{W}^{2}-s_{W}^{2})^2+4s_{W}^{4}c_{\pm}^{2})\gmn$ \\
  36        & $G^{+}G^{-}Z_{\mu}Z_{\nu}$         & $i\frac{g^{2}}{2c_{W}^{2}}(c_{\pm}^{2}(c_{W}^{2}-s_{W}^{2})^2+4s_{W}^{4}s_{\pm}^{2})\gmn$ \\
  37           & $H_{1}H_{1}Z_{\mu}Z_{\nu}$         & $i\frac{g^{2}}{2c_{W}^{2}}(\mathcal{O}^{R\,2}_{12}+4 \mathcal{O}^{R\,2}_{13})\gmn$\\
  38        & $H_{2}H_{2}Z_{\mu}Z_{\nu}$         & $i\frac{g^{2}}{2c_{W}^{2}}(\mathcal{O}^{R\,2}_{22}+4 \mathcal{O}^{R\,2}_{23})\gmn$\\
  39           & $H_{3}H_{3}Z_{\mu}Z_{\nu}$         & $i\frac{g^{2}}{2c_{W}^{2}}(\mathcal{O}^{R\,2}_{32}+4 \mathcal{O}^{R\,2}_{33})\gmn$\\
  40          &    $J J Z_{\mu}Z_{\nu}$            & $i\frac{g^{2}}{2c_{W}^{2}}(\mathcal{O}^{I\,2}_{12}+4 \mathcal{O}^{I\,2}_{13})\gmn$\\
  41        & $G^{0}G^{0}Z_{\mu}Z_{\nu}$         & $i\frac{g^{2}}{2c_{W}^{2}}(\mathcal{O}^{I\,2}_{22}+4 \mathcal{O}^{I\,2}_{23})\gmn$\\
  42           &    $A A Z_{\mu}Z_{\nu}$            & $i\frac{g^{2}}{2c_{W}^{2}}(\mathcal{O}^{I\,2}_{32}+4 \mathcal{O}^{I\,2}_{33})\gmn$\\
  43           & $\Delta^{++}\Delta^{--}Z_{\mu}$    & $-\frac{ig}{c_{W}}(c_{W}^{2}-s_{W}^{2})\ppm$ \\
  44           & $H^{-}H^{+}  Z_{\mu}$              & $\frac{ig}{2c_{W}}(s_{\pm}^{2}(c_{W}^{2}-s_{W}^{2})-2s_{W}^{2}c_{\pm}^{2})\ppm$ \\
  45        & $G^{-}G^{+}  Z_{\mu}$              & $\frac{ig}{2c_{W}}(c_{\pm}^{2}(c_{W}^{2}-s_{W}^{2})-2s_{W}^{2}s_{\pm}^{2})\ppm$ \\ 
  46           & $H_{1} J Z_{\mu}$                  & $-\frac{g}{2c_{W}}(\mathcal{O}^{R}_{12}\mathcal{O}^{I}_{12}-2\mathcal{O}^{R}_{13}\mathcal{O}^{I}_{13})\ppm$ \\ 
  47           & $G^{0}H_{1} Z_{\mu}$               & $\frac{g}{2c_{W}}(\mathcal{O}^{R}_{12}\mathcal{O}^{I}_{22}-2\mathcal{O}^{R}_{13}\mathcal{O}^{I}_{23})\ppm$ \\
  48           & $A H_{1} Z_{\mu}$                  & $\frac{g}{2c_{W}}(\mathcal{O}^{R}_{12}\mathcal{O}^{I}_{32}-2\mathcal{O}^{R}_{13}\mathcal{O}^{I}_{33})\ppm$ \\  
  49           & $H_{2} J Z_{\mu}$                  & $-\frac{g}{2c_{W}}(\mathcal{O}^{R}_{22}\mathcal{O}^{I}_{12}-2\mathcal{O}^{R}_{23}\mathcal{O}^{I}_{13})\ppm$ \\ 
  50        & $G^{0}H_{2} Z_{\mu}$               & $\frac{g}{2c_{W}}(\mathcal{O}^{R}_{22}\mathcal{O}^{I}_{22}-2\mathcal{O}^{R}_{23}\mathcal{O}^{I}_{23})\ppm$ \\
  51           & $A H_{2} Z_{\mu}$                  & $\frac{g}{2c_{W}}(\mathcal{O}^{R}_{22}\mathcal{O}^{I}_{32}-2\mathcal{O}^{R}_{23}\mathcal{O}^{I}_{33})\ppm$ \\  
  52           & $H_{3} J Z_{\mu}$                  & $-\frac{g}{2c_{W}}(\mathcal{O}^{R}_{32}\mathcal{O}^{I}_{12}-2\mathcal{O}^{R}_{33}\mathcal{O}^{I}_{13})\ppm$ \\ 
  53           & $G^{0}H_{3} Z_{\mu}$               & $\frac{g}{2c_{W}}(\mathcal{O}^{R}_{32}\mathcal{O}^{I}_{22}-2\mathcal{O}^{R}_{33}\mathcal{O}^{I}_{23})\ppm$ \\
  54           & $A H_{3} Z_{\mu}$                  & $\frac{g}{2c_{W}}(\mathcal{O}^{R}_{32}\mathcal{O}^{I}_{32}-2\mathcal{O}^{R}_{33}\mathcal{O}^{I}_{33})\ppm$ \\  
  55           & $G^{\mp}H^\pm Z_{\mu}$             & $\mp\frac{g}{2c_{W}}c_{\pm}s_{\pm}\ppm$\\
  56           & $\Delta^{++}\Delta^{--}A_{\mu}A_{\mu}$ & $8ie^{2}\gmn$\\
  57           & $H^{-}H^{+}  A_{\mu}A_{\mu}$       & $i2e^{2}\gmn$ \\ 
  58        & $G^{-}G^{+}  A_{\mu}A_{\mu}$       & $i2e^{2}\gmn$ \\ 
  59           & $\Delta^{++}\Delta^{--}A_{\mu}$    & $-2ie\ppm$\\
  60           & $H^{+}H^{-}  A_{\mu}$              & $ie\ppm$ \\ 
  61        & $G^{+}G^{-}  A_{\mu}$              & $ie\ppm$ \\ 
  62           & $\Delta^{++}\Delta^{--} A_{\mu}Z_{\nu}$ & $4i\frac{eg}{c_{W}}(c_{W}^{2}-s_{W}^{2})\gmn$ \\
  63           & $H^{+}H^{-} A_{\mu}Z_{\nu}$        & $i\frac{eg}{c_{W}}(s_{\pm}^{2}(c_{W}^{2}-s_{W}^{2})-2c_{\pm}^{2}s_{W}^{2})\gmn$ \\
  64        & $G^{+}G^{-} A_{\mu}Z_{\nu}$        & $i\frac{eg}{c_{W}}(c_{\pm}^{2}(c_{W}^{2}-s_{W}^{2})-2s_{\pm}^{2}s_{W}^{2})\gmn$ \\ 
  \hline
  \end{tabular}
  \end{tiny} 
  \caption{Feynman rules for the couplings of the Higgs bosons $H_{i}$ 
   to the gauge bosons.}
   \label{tab:2}
\end{table}

\newpage

\begin{thebibliography}{10}
\providecommand{\url}[1]{\texttt{#1}}
\providecommand{\urlprefix}{URL }
\providecommand{\eprint}[2][]{\url{#2}}

\bibitem{Aad:2012tfa}
G.~Aad et~al. (ATLAS Collaboration), \emph{{Observation of a new particle in
  the search for the Standard Model Higgs boson with the ATLAS detector at the
  LHC}},
  \MYhref[journalLinks]{http://dx.doi.org/10.1016/j.physletb.2012.08.020}{Phys.Lett.
  }\MYhref[journalLinks]{http://dx.doi.org/10.1016/j.physletb.2012.08.020}{\textbf{B716}
  (2012) 1--29},
  \MYhref[eprintLinks]{http://arxiv.org/abs/1207.7214}{{\ttfamily
  arXiv:1207.7214 [hep-ex]}}.

\bibitem{Aad:2014eva}
G.~Aad et~al. (ATLAS Collaboration), \emph{{Measurements of Higgs boson
  production and couplings in the four-lepton channel in pp collisions at
  center-of-mass energies of 7 and 8 TeV with the ATLAS detector}}  (2014),
  \MYhref[eprintLinks]{http://arxiv.org/abs/1408.5191}{{\ttfamily
  arXiv:1408.5191 [hep-ex]}}.

\bibitem{Khachatryan:2014jba}
V.~Khachatryan et~al. (CMS), \emph{{Precise determination of the mass of the
  Higgs boson and tests of compatibility of its couplings with the standard
  model predictions using proton collisions at 7 and 8 $\,\text {TeV}$}},
  \MYhref[journalLinks]{http://dx.doi.org/10.1140/epjc/s10052-015-3351-7}{Eur.
  Phys. J.
  }\MYhref[journalLinks]{http://dx.doi.org/10.1140/epjc/s10052-015-3351-7}{\textbf{C75}
  (2015) 5 212},
  \MYhref[eprintLinks]{http://arxiv.org/abs/1412.8662}{{\ttfamily
  arXiv:1412.8662 [hep-ex]}}.

\bibitem{Aad:2015gba}
G.~Aad et~al. (ATLAS), \emph{{Measurements of the Higgs boson production and
  decay rates and coupling strengths using $pp$ collision data at $\sqrt{s}=7$
  and $8$ TeV in the ATLAS experiment}}  (2015),
  \MYhref[eprintLinks]{http://arxiv.org/abs/1507.04548}{{\ttfamily
  arXiv:1507.04548 [hep-ex]}}.

\bibitem{Baak:2012kk}
M.~Baak et~al., \emph{{The Electroweak Fit of the Standard Model after the
  Discovery of a New Boson at the LHC}},
  \MYhref[journalLinks]{http://dx.doi.org/10.1140/epjc/s10052-012-2205-9}{Eur.
  Phys. J.
  }\MYhref[journalLinks]{http://dx.doi.org/10.1140/epjc/s10052-012-2205-9}{\textbf{C72}
  (2012) 2205}, \MYhref[eprintLinks]{http://arxiv.org/abs/1209.2716}{{\ttfamily
  arXiv:1209.2716 [hep-ph]}}.

\bibitem{Baak:2014ora}
M.~Baak et~al. (Gfitter Group), \emph{{The global electroweak fit at NNLO and
  prospects for the LHC and ILC}},
  \MYhref[journalLinks]{http://dx.doi.org/10.1140/epjc/s10052-014-3046-5}{Eur.
  Phys. J.
  }\MYhref[journalLinks]{http://dx.doi.org/10.1140/epjc/s10052-014-3046-5}{\textbf{C74}
  (2014) 3046}, \MYhref[eprintLinks]{http://arxiv.org/abs/1407.3792}{{\ttfamily
  arXiv:1407.3792 [hep-ph]}}.

\bibitem{PhysRevLett.65.964}
M.~E. Peskin and T.~Takeuchi, \emph{New constraint on a strongly interacting
  higgs sector},
  \MYhref[journalLinks]{http://dx.doi.org/10.1103/PhysRevLett.65.964}{Phys.
  Rev. Lett.
  }\MYhref[journalLinks]{http://dx.doi.org/10.1103/PhysRevLett.65.964}{\textbf{65}
  (1990) 964--967},
  \urlprefix\url{http://link.aps.org/doi/10.1103/PhysRevLett.65.964}.

\bibitem{ALTARELLI1991161}
G.~Altarelli and R.~Barbieri, \emph{Vacuum polarization effects of new physics
  on electroweak processes},
  \MYhref[journalLinks]{http://dx.doi.org/http://dx.doi.org/10.1016/0370-2693(91)91378-9}{Physics
  Letters B
  }\MYhref[journalLinks]{http://dx.doi.org/http://dx.doi.org/10.1016/0370-2693(91)91378-9}{\textbf{253}
  (1991) 1 161 -- 167}, ISSN 0370-2693,
  \urlprefix\url{http://www.sciencedirect.com/science/article/pii/0370269391913789}.

\bibitem{PhysRevD.46.381}
M.~E. Peskin and T.~Takeuchi, \emph{Estimation of oblique electroweak
  corrections},
  \MYhref[journalLinks]{http://dx.doi.org/10.1103/PhysRevD.46.381}{Phys. Rev. D
  }\MYhref[journalLinks]{http://dx.doi.org/10.1103/PhysRevD.46.381}{\textbf{46}
  (1992) 381--409},
  \urlprefix\url{http://link.aps.org/doi/10.1103/PhysRevD.46.381}.

\bibitem{Djouadi:2005gi}
A.~Djouadi, \emph{{The Anatomy of electro-weak symmetry breaking. I: The Higgs
  boson in the standard model}}, Phys.Rept. \textbf{457} (2008) 1--216,
  \MYhref[eprintLinks]{http://arxiv.org/abs/hep-ph/0503172}{{\ttfamily
  arXiv:hep-ph/0503172 [hep-ph]}}.

\bibitem{Bonilla:2015kna}
C.~Bonilla, R.~M. Fonseca and J.~W.~F. Valle, \emph{{Vacuum stability with
  spontaneous violation of lepton number}}  (2015),
  \MYhref[eprintLinks]{http://arxiv.org/abs/1506.04031}{{\ttfamily
  arXiv:1506.04031 [hep-ph]}}.

\bibitem{Bonilla:2015eha}
C.~Bonilla, R.~M. Fonseca and J.~W.~F. Valle, \emph{{Consistency of the triplet
  seesaw model revisited}},
  \MYhref[journalLinks]{http://dx.doi.org/10.1103/PhysRevD.92.075028}{Phys.
  Rev.
  }\MYhref[journalLinks]{http://dx.doi.org/10.1103/PhysRevD.92.075028}{\textbf{D92}
  (2015) 7 075028},
  \MYhref[eprintLinks]{http://arxiv.org/abs/1508.02323}{{\ttfamily
  arXiv:1508.02323 [hep-ph]}}.

\bibitem{Bonilla:2015uwa}
C.~Bonilla, J.~W.~F. Valle and J.~C. Romão, \emph{{Neutrino mass and invisible
  Higgs decays at the LHC}},
  \MYhref[journalLinks]{http://dx.doi.org/10.1103/PhysRevD.91.113015}{Phys.
  Rev.
  }\MYhref[journalLinks]{http://dx.doi.org/10.1103/PhysRevD.91.113015}{\textbf{D91}
  (2015) 11 113015},
  \MYhref[eprintLinks]{http://arxiv.org/abs/1502.01649}{{\ttfamily
  arXiv:1502.01649 [hep-ph]}}.

\bibitem{Schechter:1981cv}
J.~Schechter and J.~Valle, \emph{{Neutrino Decay and Spontaneous Violation of
  Lepton Number}},
  \MYhref[journalLinks]{http://dx.doi.org/10.1103/PhysRevD.25.774}{Phys.Rev.
  }\MYhref[journalLinks]{http://dx.doi.org/10.1103/PhysRevD.25.774}{\textbf{D25}
  (1982) 774}.

\bibitem{Joshipura:1992hp}
A.~S. Joshipura and J.~Valle, \emph{{Invisible Higgs decays and neutrino
  physics}},
  \MYhref[journalLinks]{http://dx.doi.org/10.1016/0550-3213(93)90337-O}{Nucl.Phys.
  }\MYhref[journalLinks]{http://dx.doi.org/10.1016/0550-3213(93)90337-O}{\textbf{B397}
  (1993) 105--122}.

\bibitem{Diaz:1998zg}
M.~A. Diaz, M.~Garcia-Jareno, D.~A. Restrepo and J.~Valle, \emph{{Seesaw
  Majoron model of neutrino mass and novel signals in Higgs boson production at
  LEP}},
  \MYhref[journalLinks]{http://dx.doi.org/10.1016/S0550-3213(98)00434-9}{Nucl.Phys.
  }\MYhref[journalLinks]{http://dx.doi.org/10.1016/S0550-3213(98)00434-9}{\textbf{B527}
  (1998) 44--60},
  \MYhref[eprintLinks]{http://arxiv.org/abs/hep-ph/9803362}{{\ttfamily
  arXiv:hep-ph/9803362 [hep-ph]}}.

\bibitem{Chikashige:1980ui}
Y.~Chikashige, R.~N. Mohapatra and R.~Peccei, \emph{{Are There Real Goldstone
  Bosons Associated with Broken Lepton Number?}},
  \MYhref[journalLinks]{http://dx.doi.org/10.1016/0370-2693(81)90011-3}{Phys.Lett.
  }\MYhref[journalLinks]{http://dx.doi.org/10.1016/0370-2693(81)90011-3}{\textbf{B98}
  (1981) 265}.

\bibitem{Gelmini:1980re}
G.~Gelmini and M.~Roncadelli, \emph{{Left-Handed Neutrino Mass Scale and
  Spontaneously Broken Lepton Number}},
  \MYhref[journalLinks]{http://dx.doi.org/10.1016/0370-2693(81)90559-1}{Phys.Lett.
  }\MYhref[journalLinks]{http://dx.doi.org/10.1016/0370-2693(81)90559-1}{\textbf{B99}
  (1981) 411}.

\bibitem{Shrock:1982kd}
R.~E. Shrock and M.~Suzuki, \emph{{Invisible Decays of Higgs Bosons}},
  \MYhref[journalLinks]{http://dx.doi.org/10.1016/0370-2693(82)91247-3}{Phys.
  Lett.
  }\MYhref[journalLinks]{http://dx.doi.org/10.1016/0370-2693(82)91247-3}{\textbf{B110}
  (1982) 250}.

\bibitem{Boucenna:2014zba}
S.~M. Boucenna, S.~Morisi and J.~W. Valle, \emph{{The low-scale approach to
  neutrino masses}},
  \MYhref[journalLinks]{http://dx.doi.org/10.1155/2014/831598}{Adv.High Energy
  Phys.
  }\MYhref[journalLinks]{http://dx.doi.org/10.1155/2014/831598}{\textbf{2014}
  (2014) 831598},
  \MYhref[eprintLinks]{http://arxiv.org/abs/1404.3751}{{\ttfamily
  arXiv:1404.3751 [hep-ph]}}.

\bibitem{Schechter:1980gr}
J.~Schechter and J.~Valle, \emph{{Neutrino Masses in SU(2) x U(1) Theories}},
  \MYhref[journalLinks]{http://dx.doi.org/10.1103/PhysRevD.22.2227}{Phys.Rev.
  }\MYhref[journalLinks]{http://dx.doi.org/10.1103/PhysRevD.22.2227}{\textbf{D22}
  (1980) 2227}.

\bibitem{Kannike:2012pe}
K.~Kannike, \emph{{Vacuum Stability Conditions From Copositivity Criteria}},
  \MYhref[journalLinks]{http://dx.doi.org/10.1140/epjc/s10052-012-2093-z}{Eur.Phys.J.
  }\MYhref[journalLinks]{http://dx.doi.org/10.1140/epjc/s10052-012-2093-z}{\textbf{C72}
  (2012) 2093}, \MYhref[eprintLinks]{http://arxiv.org/abs/1205.3781}{{\ttfamily
  arXiv:1205.3781 [hep-ph]}}.

\bibitem{Agashe:2014kda}
K.~A. Olive et~al. (Particle Data Group), \emph{{Review of Particle Physics}},
  \MYhref[journalLinks]{http://dx.doi.org/10.1088/1674-1137/38/9/090001}{Chin.
  Phys.
  }\MYhref[journalLinks]{http://dx.doi.org/10.1088/1674-1137/38/9/090001}{\textbf{C38}
  (2014) 090001}.

\bibitem{Choi:1989hi}
K.~Choi and A.~Santamaria, \emph{{Majorons and Supernova Cooling}},
  \MYhref[journalLinks]{http://dx.doi.org/10.1103/PhysRevD.42.293}{Phys.Rev.
  }\MYhref[journalLinks]{http://dx.doi.org/10.1103/PhysRevD.42.293}{\textbf{D42}
  (1990) 293--306}.

\bibitem{Khachatryan:2014ira}
V.~Khachatryan et~al. (CMS Collaboration), \emph{{Observation of the diphoton
  decay of the Higgs boson and measurement of its properties}},
  \MYhref[journalLinks]{http://dx.doi.org/10.1140/epjc/s10052-014-3076-z}{Eur.Phys.J.
  }\MYhref[journalLinks]{http://dx.doi.org/10.1140/epjc/s10052-014-3076-z}{\textbf{C74}
  (2014) 10 3076},
  \MYhref[eprintLinks]{http://arxiv.org/abs/1407.0558}{{\ttfamily
  arXiv:1407.0558 [hep-ex]}}.

\bibitem{Aad:2015zhl}
G.~Aad et~al. (ATLAS, CMS), \emph{{Combined Measurement of the Higgs Boson Mass
  in $pp$ Collisions at $\sqrt{s}=7$ and 8 TeV with the ATLAS and CMS
  Experiments}},
  \MYhref[journalLinks]{http://dx.doi.org/10.1103/PhysRevLett.114.191803}{Phys.
  Rev. Lett.
  }\MYhref[journalLinks]{http://dx.doi.org/10.1103/PhysRevLett.114.191803}{\textbf{114}
  (2015) 191803},
  \MYhref[eprintLinks]{http://arxiv.org/abs/1503.07589}{{\ttfamily
  arXiv:1503.07589 [hep-ex]}}.

\bibitem{Abdallah:2004wy}
J.~Abdallah et~al. (DELPHI Collaboration), \emph{{Searches for neutral higgs
  bosons in extended models}},
  \MYhref[journalLinks]{http://dx.doi.org/10.1140/epjc/s2004-02011-4}{Eur.Phys.J.
  }\MYhref[journalLinks]{http://dx.doi.org/10.1140/epjc/s2004-02011-4}{\textbf{C38}
  (2004) 1--28},
  \MYhref[eprintLinks]{http://arxiv.org/abs/hep-ex/0410017}{{\ttfamily
  arXiv:hep-ex/0410017 [hep-ex]}}.

\bibitem{ATLAS-CONF-2015-044}
\emph{{Measurements of the Higgs boson production and decay rates and
  constraints on its couplings from a combined ATLAS and CMS analysis of the
  LHC pp collision data at $\sqrt{s}$ = 7 and 8 TeV}}, Technical Report
  ATLAS-CONF-2015-044, CERN, Geneva (2015),
  \urlprefix\url{http://cds.cern.ch/record/2052552}.

\bibitem{Khachatryan:2015cwa}
V.~Khachatryan et~al. (CMS), \emph{{Search for a Higgs Boson in the Mass Range
  from 145 to 1000 GeV Decaying to a Pair of W or Z Bosons}}  (2015),
  \MYhref[eprintLinks]{http://arxiv.org/abs/1504.00936}{{\ttfamily
  arXiv:1504.00936 [hep-ex]}}.

\bibitem{Khachatryan:2014wca}
V.~Khachatryan et~al. (CMS), \emph{{Search for neutral MSSM Higgs bosons
  decaying to a pair of tau leptons in pp collisions}},
  \MYhref[journalLinks]{http://dx.doi.org/10.1007/JHEP10(2014)160}{JHEP
  }\MYhref[journalLinks]{http://dx.doi.org/10.1007/JHEP10(2014)160}{\textbf{10}
  (2014) 160}, \MYhref[eprintLinks]{http://arxiv.org/abs/1408.3316}{{\ttfamily
  arXiv:1408.3316 [hep-ex]}}.

\bibitem{Aad:2014ioa}
G.~Aad et~al. (ATLAS), \emph{{Search for Scalar Diphoton Resonances in the Mass
  Range $65-600$ GeV with the ATLAS Detector in $pp$ Collision Data at
  $\sqrt{s}$ = 8 $TeV$}},
  \MYhref[journalLinks]{http://dx.doi.org/10.1103/PhysRevLett.113.171801}{Phys.
  Rev. Lett.
  }\MYhref[journalLinks]{http://dx.doi.org/10.1103/PhysRevLett.113.171801}{\textbf{113}
  (2014) 17 171801},
  \MYhref[eprintLinks]{http://arxiv.org/abs/1407.6583}{{\ttfamily
  arXiv:1407.6583 [hep-ex]}}.

\bibitem{Khachatryan:2015qba}
V.~Khachatryan et~al. (CMS), \emph{{Search for diphoton resonances in the mass
  range from 150 to 850 GeV in pp collisions at $\sqrt{s} =$ 8 TeV}},
  \MYhref[journalLinks]{http://dx.doi.org/10.1016/j.physletb.2015.09.062}{Phys.
  Lett.
  }\MYhref[journalLinks]{http://dx.doi.org/10.1016/j.physletb.2015.09.062}{\textbf{B750}
  (2015) 494--519},
  \MYhref[eprintLinks]{http://arxiv.org/abs/1506.02301}{{\ttfamily
  arXiv:1506.02301 [hep-ex]}}.

\bibitem{Aad:2015wra}
G.~Aad et~al. (ATLAS), \emph{{Search for a CP-odd Higgs boson decaying to Zh in
  pp collisions at $\sqrt{s} = 8$ TeV with the ATLAS detector}},
  \MYhref[journalLinks]{http://dx.doi.org/10.1016/j.physletb.2015.03.054}{Phys.
  Lett.
  }\MYhref[journalLinks]{http://dx.doi.org/10.1016/j.physletb.2015.03.054}{\textbf{B744}
  (2015) 163--183},
  \MYhref[eprintLinks]{http://arxiv.org/abs/1502.04478}{{\ttfamily
  arXiv:1502.04478 [hep-ex]}}.

\bibitem{Garayoa:2007fw}
J.~Garayoa and T.~Schwetz, \emph{{Neutrino mass hierarchy and Majorana CP
  phases within the Higgs triplet model at the LHC}},
  \MYhref[journalLinks]{http://dx.doi.org/10.1088/1126-6708/2008/03/009}{JHEP
  }\MYhref[journalLinks]{http://dx.doi.org/10.1088/1126-6708/2008/03/009}{\textbf{03}
  (2008) 009}, \MYhref[eprintLinks]{http://arxiv.org/abs/0712.1453}{{\ttfamily
  arXiv:0712.1453 [hep-ph]}}.

\bibitem{Perez:2008zc}
P.~Fileviez~Perez et~al., \emph{{Testing a Neutrino Mass Generation Mechanism
  at the LHC}},
  \MYhref[journalLinks]{http://dx.doi.org/10.1103/PhysRevD.78.071301}{Phys.Rev.
  }\MYhref[journalLinks]{http://dx.doi.org/10.1103/PhysRevD.78.071301}{\textbf{D78}
  (2008) 071301},
  \MYhref[eprintLinks]{http://arxiv.org/abs/0803.3450}{{\ttfamily
  arXiv:0803.3450 [hep-ph]}}.

\bibitem{Perez:2008ha}
P.~Fileviez~Perez et~al., \emph{{Neutrino Masses and the CERN LHC: Testing Type
  II Seesaw}},
  \MYhref[journalLinks]{http://dx.doi.org/10.1103/PhysRevD.78.015018}{Phys.Rev.
  }\MYhref[journalLinks]{http://dx.doi.org/10.1103/PhysRevD.78.015018}{\textbf{D78}
  (2008) 015018},
  \MYhref[eprintLinks]{http://arxiv.org/abs/0805.3536}{{\ttfamily
  arXiv:0805.3536 [hep-ph]}}.

\bibitem{delAguila:2008cj}
F.~del Aguila and J.~A. Aguilar-Saavedra, \emph{{Distinguishing seesaw models
  at LHC with multi-lepton signals}},
  \MYhref[journalLinks]{http://dx.doi.org/10.1016/j.nuclphysb.2008.12.029}{Nucl.
  Phys.
  }\MYhref[journalLinks]{http://dx.doi.org/10.1016/j.nuclphysb.2008.12.029}{\textbf{B813}
  (2009) 22--90},
  \MYhref[eprintLinks]{http://arxiv.org/abs/0808.2468}{{\ttfamily
  arXiv:0808.2468 [hep-ph]}}.

\bibitem{Aoki:2011pz}
M.~Aoki, S.~Kanemura and K.~Yagyu, \emph{{Testing the Higgs triplet model with
  the mass difference at the LHC}},
  \MYhref[journalLinks]{http://dx.doi.org/10.1103/PhysRevD.85.055007}{Phys.
  Rev.
  }\MYhref[journalLinks]{http://dx.doi.org/10.1103/PhysRevD.85.055007}{\textbf{D85}
  (2012) 055007},
  \MYhref[eprintLinks]{http://arxiv.org/abs/1110.4625}{{\ttfamily
  arXiv:1110.4625 [hep-ph]}}.

\bibitem{Akeroyd:2012ms}
A.~G. Akeroyd and S.~Moretti, \emph{{Enhancement of H to gamma gamma from
  doubly charged scalars in the Higgs Triplet Model}},
  \MYhref[journalLinks]{http://dx.doi.org/10.1103/PhysRevD.86.035015}{Phys.
  Rev.
  }\MYhref[journalLinks]{http://dx.doi.org/10.1103/PhysRevD.86.035015}{\textbf{D86}
  (2012) 035015},
  \MYhref[eprintLinks]{http://arxiv.org/abs/1206.0535}{{\ttfamily
  arXiv:1206.0535 [hep-ph]}}.

\bibitem{Dev:2013ff}
P.~S. Bhupal~Dev, D.~K. Ghosh, N.~Okada and I.~Saha, \emph{{125 GeV Higgs Boson
  and the Type-II Seesaw Model}},
  \MYhref[journalLinks]{http://dx.doi.org/10.1007/JHEP03(2013)150,
  10.1007/JHEP05(2013)049}{JHEP
  }\MYhref[journalLinks]{http://dx.doi.org/10.1007/JHEP03(2013)150,
  10.1007/JHEP05(2013)049}{\textbf{03} (2013) 150}, [Erratum:
  JHEP05,049(2013)],
  \MYhref[eprintLinks]{http://arxiv.org/abs/1301.3453}{{\ttfamily
  arXiv:1301.3453 [hep-ph]}}.

\bibitem{delAguila:2013mia}
F.~del \'{A}guila and M.~Chala, \emph{{LHC bounds on Lepton Number Violation
  mediated by doubly and singly-charged scalars}},
  \MYhref[journalLinks]{http://dx.doi.org/10.1007/JHEP03(2014)027}{JHEP
  }\MYhref[journalLinks]{http://dx.doi.org/10.1007/JHEP03(2014)027}{\textbf{03}
  (2014) 027}, \MYhref[eprintLinks]{http://arxiv.org/abs/1311.1510}{{\ttfamily
  arXiv:1311.1510 [hep-ph]}}.

\bibitem{Chen:2014qda}
C.-H. Chen and T.~Nomura, \emph{{Search for $\delta^{\pm\pm}$ with new decay
  patterns at the LHC}},
  \MYhref[journalLinks]{http://dx.doi.org/10.1103/PhysRevD.91.035023}{Phys.
  Rev.
  }\MYhref[journalLinks]{http://dx.doi.org/10.1103/PhysRevD.91.035023}{\textbf{D91}
  (2015) 035023},
  \MYhref[eprintLinks]{http://arxiv.org/abs/1411.6412}{{\ttfamily
  arXiv:1411.6412 [hep-ph]}}.

\bibitem{Kanemura:2014goa}
S.~Kanemura, M.~Kikuchi, K.~Yagyu and H.~Yokoya, \emph{{Bounds on the mass of
  doubly-charged Higgs bosons in the same-sign diboson decay scenario}},
  \MYhref[journalLinks]{http://dx.doi.org/10.1103/PhysRevD.90.115018}{Phys.
  Rev.
  }\MYhref[journalLinks]{http://dx.doi.org/10.1103/PhysRevD.90.115018}{\textbf{D90}
  (2014) 11 115018},
  \MYhref[eprintLinks]{http://arxiv.org/abs/1407.6547}{{\ttfamily
  arXiv:1407.6547 [hep-ph]}}.

\bibitem{Han:2015hba}
Z.-L. Han, R.~Ding and Y.~Liao, \emph{{LHC Phenomenology of Type II Seesaw:
  Nondegenerate Case}},
  \MYhref[journalLinks]{http://dx.doi.org/10.1103/PhysRevD.91.093006}{Phys.
  Rev.
  }\MYhref[journalLinks]{http://dx.doi.org/10.1103/PhysRevD.91.093006}{\textbf{D91}
  (2015) 093006},
  \MYhref[eprintLinks]{http://arxiv.org/abs/1502.05242}{{\ttfamily
  arXiv:1502.05242 [hep-ph]}}.

\bibitem{Sugiyama:2012yw}
H.~Sugiyama, K.~Tsumura and H.~Yokoya, \emph{{Discrimination of models
  including doubly charged scalar bosons by using tau lepton decay
  distributions}},
  \MYhref[journalLinks]{http://dx.doi.org/10.1016/j.physletb.2012.09.044}{Phys.
  Lett.
  }\MYhref[journalLinks]{http://dx.doi.org/10.1016/j.physletb.2012.09.044}{\textbf{B717}
  (2014) 229--234},
  \MYhref[eprintLinks]{http://arxiv.org/abs/1207.0179}{{\ttfamily
  arXiv:1207.0179 [hep-ph]}}.

\bibitem{Chatrchyan:2012ya}
S.~Chatrchyan et~al. (CMS), \emph{{A search for a doubly-charged Higgs boson in
  $pp$ collisions at $\sqrt{s}=7$ TeV}},
  \MYhref[journalLinks]{http://dx.doi.org/10.1140/epjc/s10052-012-2189-5}{Eur.
  Phys. J.
  }\MYhref[journalLinks]{http://dx.doi.org/10.1140/epjc/s10052-012-2189-5}{\textbf{C72}
  (2012) 2189}, \MYhref[eprintLinks]{http://arxiv.org/abs/1207.2666}{{\ttfamily
  arXiv:1207.2666 [hep-ex]}}.

\bibitem{ATLAS:2012hi}
G.~Aad et~al. (ATLAS), \emph{{Search for doubly-charged Higgs bosons in
  like-sign dilepton final states at $\sqrt{s}=7$ TeV with the ATLAS
  detector}},
  \MYhref[journalLinks]{http://dx.doi.org/10.1140/epjc/s10052-012-2244-2}{Eur.
  Phys. J.
  }\MYhref[journalLinks]{http://dx.doi.org/10.1140/epjc/s10052-012-2244-2}{\textbf{C72}
  (2012) 2244}, \MYhref[eprintLinks]{http://arxiv.org/abs/1210.5070}{{\ttfamily
  arXiv:1210.5070 [hep-ex]}}.

\bibitem{Kanemura:2013vxa}
S.~Kanemura, K.~Yagyu and H.~Yokoya, \emph{{First constraint on the mass of
  doubly-charged Higgs bosons in the same-sign diboson decay scenario at the
  LHC}},
  \MYhref[journalLinks]{http://dx.doi.org/10.1016/j.physletb.2013.08.054}{Phys.
  Lett.
  }\MYhref[journalLinks]{http://dx.doi.org/10.1016/j.physletb.2013.08.054}{\textbf{B726}
  (2013) 316--319},
  \MYhref[eprintLinks]{http://arxiv.org/abs/1305.2383}{{\ttfamily
  arXiv:1305.2383 [hep-ph]}}.

\bibitem{Kanemura:2014ipa}
S.~Kanemura, M.~Kikuchi, H.~Yokoya and K.~Yagyu, \emph{{LHC Run-I constraint on
  the mass of doubly charged Higgs bosons in the same-sign diboson decay
  scenario}}, \MYhref[journalLinks]{http://dx.doi.org/10.1093/ptep/ptv071}{PTEP
  }\MYhref[journalLinks]{http://dx.doi.org/10.1093/ptep/ptv071}{\textbf{2015}
  (2015) 051B02},
  \MYhref[eprintLinks]{http://arxiv.org/abs/1412.7603}{{\ttfamily
  arXiv:1412.7603 [hep-ph]}}.

\bibitem{Khachatryan:2014sta}
V.~Khachatryan et~al. (CMS), \emph{{Study of vector boson scattering and search
  for new physics in events with two same-sign leptons and two jets}},
  \MYhref[journalLinks]{http://dx.doi.org/10.1103/PhysRevLett.114.051801}{Phys.
  Rev. Lett.
  }\MYhref[journalLinks]{http://dx.doi.org/10.1103/PhysRevLett.114.051801}{\textbf{114}
  (2015) 5 051801},
  \MYhref[eprintLinks]{http://arxiv.org/abs/1410.6315}{{\ttfamily
  arXiv:1410.6315 [hep-ex]}}.

\bibitem{Fontes:2014xva}
D.~Fontes, J.~C. Romão and J.~P. Silva, \emph{{$h \rightarrow Z \gamma$ in the
  complex two Higgs doublet model}},
  \MYhref[journalLinks]{http://dx.doi.org/10.1007/JHEP12(2014)043}{JHEP
  }\MYhref[journalLinks]{http://dx.doi.org/10.1007/JHEP12(2014)043}{\textbf{12}
  (2014) 043}, \MYhref[eprintLinks]{http://arxiv.org/abs/1408.2534}{{\ttfamily
  arXiv:1408.2534 [hep-ph]}}.

\bibitem{Aad:2015txa}
G.~Aad et~al. (ATLAS), \emph{{Search for invisible decays of a Higgs boson
  using vector-boson fusion in $pp$ collisions at $\sqrt{s}=8$ TeV with the
  ATLAS detector}}  (2015),
  \MYhref[eprintLinks]{http://arxiv.org/abs/1508.07869}{{\ttfamily
  arXiv:1508.07869 [hep-ex]}}.

\bibitem{Chatrchyan:2014tja}
S.~Chatrchyan et~al. (CMS), \emph{{Search for invisible decays of Higgs bosons
  in the vector boson fusion and associated ZH production modes}},
  \MYhref[journalLinks]{http://dx.doi.org/10.1140/epjc/s10052-014-2980-6}{Eur.
  Phys. J.
  }\MYhref[journalLinks]{http://dx.doi.org/10.1140/epjc/s10052-014-2980-6}{\textbf{C74}
  (2014) 2980}, \MYhref[eprintLinks]{http://arxiv.org/abs/1404.1344}{{\ttfamily
  arXiv:1404.1344 [hep-ex]}}.

\bibitem{Passarino:1978jh}
G.~Passarino and M.~J.~G. Veltman, \emph{{One loop corrections for $e^+ e^-$
  annihilation into $\mu^+ \mu^-$ in the Weinberg model}},
  \MYhref[journalLinks]{http://dx.doi.org/http://dx.doi.org/10.1016/0550-3213(79)90234-7}{Nucl.
  Phys.
  }\MYhref[journalLinks]{http://dx.doi.org/http://dx.doi.org/10.1016/0550-3213(79)90234-7}{\textbf{B160}
  (1979) 151--207},
  \urlprefix\url{http://www.sciencedirect.com/science/article/pii/0550321379902347}.

\end{thebibliography}

\end{document}